\documentclass[letterpaper,journal]{IEEEtran}
\usepackage{amsmath,amsfonts}
\usepackage{algorithmic}
\usepackage{algorithm}
\usepackage{array}
\usepackage[caption=false,font=normalsize,labelfont=sf,textfont=sf]{subfig}
\usepackage{textcomp}
\usepackage{stfloats}
\usepackage{url}
\usepackage{verbatim}
\usepackage{graphicx}
\usepackage{epstopdf}
\usepackage{cite}

\newtheorem{proposition}{Proposition}

\newtheorem{remark}{Remark}

\usepackage[T1]{fontenc}
\usepackage[utf8]{inputenc}
\usepackage{amssymb,bm,booktabs}
\usepackage[hidelinks]{hyperref}
\newcommand{\papertableformat}{%
  \small
  \renewcommand{\arraystretch}{1.15}%
  \setlength{\tabcolsep}{3pt}%
}
\begin{document}
\raggedbottom

\title{Geometry–Time Graph Attention for 3-D TOA Localization in Underwater Acoustic Sensor Networks}

% TODO: Add the actual authors, affiliations, and acknowledgments.
\author{Wei Huang,~\IEEEmembership{Member,~IEEE,} Hongfei Chang, Kaitao Meng,~\IEEEmembership{Senior Member,~IEEE,} Tianhe Xu, Hao Zhang,~\IEEEmembership{Senior Member,~IEEE,} Yanan Wu, Qian Sun
	% <-this % stops a space
	\thanks{This work was supported in part by the National Key Research and Development Program of China under Grant 2024YFB3909701, in part by the National Natural Science Foundation of China under Grants 42404001 and 62271459, in part by Shandong Province Natural Science Foundation under Grants ZR2023QF128 and ZR2024QF014, in part by Central University Basic Research Fund of China, Ocean University of China under Grant 202562022.}% <-this % stops a space
	\thanks{Wei Huang, Hongfei Chang, Hao Zhang and Yanan Wu are with the Faculty of Information Science and Engineering, Ocean University of China, Qingdao, Shandong 266100, China (email: hw@ouc.edu.cn; hongfeichang@stu.ouc.edu.cn; qq258357656@163.com; wuyanan@ouc.edu.cn). Kaitao Meng is with the Electronic Information School, Wuhan University, Wuhan, China (kaitao.meng@whu.edu.cn). Tianhe Xu is with the Shandong University (Weihai), Weihai, 264200, China (email: thxu@sdu.edu.cn). Qian Sun is with the China Waterborne Transport Research Institute, Beijing, China (qbcouple@163.com).}
	\thanks{Corresponding author: Kaitao Meng (email: kaitao.meng@whu.edu.cn).}
}
\maketitle

\begin{abstract}
In underwater acoustic sensor networks, depth-dependent variations in sound speed refract acoustic rays, complicating the estimation of target positions from travel-time measurements. Propagation-model-based methods rely on an assumed or estimated sound speed profile (SSP), and a mismatch between that profile and the current environment can bias position estimates.
To address this limitation, this paper proposes an underwater graph attention network (UGAN) that fuses synchronized time-of-arrival (TOA) observations from multiple stations with reference-station coordinates to estimate three-dimensional target positions without requiring the current sound speed profile at inference. UGAN constructs one graph per target for batch inference. Its geometry--time attention mechanism combines interstation geometry with ordered pairs of TOA observations to compute attention weights in all three graph layers and enrich the
messages in the last two layers. A shared nonlinear encoder supplies individual TOA features directly to every layer. In ray-tracing simulations with a fixed five-station deployment and two SSPs held out from training and validation, UGAN achieves mean localization root mean square errors (RMSE) of 0.307, 0.577, and 0.989~m at three increasing TOA noise levels, representing reductions of 52.0\%, 31.0\%, and 22.8\% relative to the evaluated GATv2 baseline, respectively. At the intermediate noise level, incorporating station-pair relations into the messages reduces the mean localization RMSE by 9.4\% with relation-based attention scoring enabled. For 4,000 target observations, UGAN takes a median of 58.14~ms on the same GPU, 7.0\% less time than GATv2.
\end{abstract}


\section{Introduction}
\label{sec:1}
Accurate localization supports ocean observation and the navigation of autonomous underwater vehicles. Position estimates allow ocean measurements to be associated with their collection locations and provide vehicles with the position updates needed for navigation. In underwater acoustic sensor networks, the long propagation range of acoustic signals allows spatially distributed reference stations to observe the same target \cite{akyildiz,tan,erol}. Travel-time measurements from these stations provide complementary constraints on the target position.  The accuracy of the resulting position estimate depends on the reference-station geometry and the acoustic propagation environment \cite{paull,toky}.

In a stratified ocean, variations in temperature, salinity, and pressure with depth produce sound-speed gradients that refract acoustic rays. A constant-speed, straight-path approximation can therefore introduce systematic errors when travel times are converted to ranges. Stratification compensation and ray-tracing methods account for refraction using a sound speed profile (SSP) \cite{berger,rayls}. Their travel-time predictions are sensitive to how well the assumed SSP represents the current propagation environment \cite{cheniso}. Errors in reference-station coordinates also perturb the assumed propagation endpoints, while the station geometry affects how travel-time errors translate into position errors \cite{calibration,geometry}. This work considers three-dimensional localization from multistation travel-time observations and reference-station coordinates when the current SSP is unavailable at inference.

When the SSP is uncertain, model-based methods can jointly estimate the target position and the parameters of an assumed SSP. Their accuracy depends on how well the assumed profile, such as an isogradient SSP, represents the propagation environment \cite{cheniso}. Data-driven methods learn to estimate source range or position from acoustic observations \cite{bianco}. Studies using array-pressure features have explored simulation-based training, transfer to experimental data, and training across diverse acoustic environments \cite{niu,wangtransfer,chencnn,diniz}. Learned estimators have also been combined with stratification compensation for underwater sensor networks \cite{broad}. The performance of these estimators depends on the environments covered by the training data and on how the observations are represented. For multistation time-of-arrival (TOA) localization, a key design problem is how to fuse timing observations from spatially distributed reference stations while accounting for their relative geometry.

A graph provides a natural representation for multistation TOA localization, with reference stations as nodes and station pairs as edges \cite{bronstein,wu}. The same difference in travel time can impose different position constraints for station pairs with different horizontal separations or depth differences. This motivates representing paired TOA observations together with the corresponding station geometry and using the resulting station-pair features to inform both attention weights and message content. Each individual TOA also provides timing information for its own target--station link. Giving each graph layer direct access to individual TOA features allows it to use this link-level information alongside features aggregated in earlier layers.

To fuse multistation TOA observations when the current SSP is unavailable at inference, we propose an Underwater Graph Attention Network (UGAN) for three-dimensional localization. UGAN constructs a separate graph for each target from its TOA observations and the corresponding reference-station coordinates. Its Geometry--Time Attention (GTA) mechanism uses station-pair geometry and paired TOA observations to guide aggregation, while a separate encoder gives each graph layer direct access to individual TOA features. The main contributions of this integrated design are as follows.
\begin{itemize}
\item To support localization of multiple targets from distributed reference-station measurements, we develop UGAN, a graph-based framework that maps each target-specific observation graph to a three-dimensional position. The same trained model processes multiple graphs in a batch without requiring the current SSP.

\item To account for station geometry when interpreting paired TOA observations, we jointly encode relative station displacements, interstation distances, and ordered TOA pairs. The resulting station-pair features contribute to dynamic attention scoring in all three graph layers and to message aggregation in the final two layers.

\item To give each graph layer direct access to individual TOA observations, we use a nonlinear encoder shared across reference stations and separate from the paired-TOA encoders. Layer-specific projections inject the individual TOA features into the node messages of all three layers.

\item To assess localization accuracy and execution efficiency, we evaluate UGAN at three TOA noise levels using two held-out SSPs and a fixed five-station deployment. UGAN reduces the mean
localization root mean square error (RMSE) by $22.8\%$ to $52.0\%$ relative to GATv2~\cite{gatv2}. For a workload of 4,000 target observations, the median processing time on the same GPU is
$58.14$~ms for UGAN and $62.51$~ms for GATv2, a reduction of $7.0\%$.
\end{itemize}

The remainder of this paper is organized as follows. Section~\ref{sec:2} reviews related work, and Section~\ref{sec:3} presents the observation model, conditional Cramér--Rao lower bound, and UGAN. Section~\ref{sec:4} describes the simulation setup and localization results. Section~\ref{sec:5} concludes the paper and discusses limitations and future work.

\section{Related Work}
\label{sec:2}

\subsection{Propagation-Model-Based Localization and Error Sources}
\label{sec:2.1}
Propagation-model-based localization relates acoustic measurements to the position of a target through a model of signal travel time. The resulting position constraints depend on the measurement type, synchronization conditions, and reference-station deployment \cite{tan,erol,toky}. When clock offsets are unknown, the observation model must account for their effects on measured arrival times. Approaches for mobile underwater networks address this issue through joint synchronization and localization \cite{jointsync} or through mobility prediction in asynchronous localization \cite{async}. Under the synchronized TOA setting considered here, the main modeling question is how to relate the resulting travel-time observations to target position.

The simplest approach converts travel times to ranges using a constant sound speed and straight-line paths. In a stratified ocean, depth-dependent sound speed bends acoustic rays and changes the relationship between travel time and range. Stratification compensation addresses this effect in acoustic ranging \cite{berger}, while ray-tracing localization predicts travel times along curved paths and estimates position by comparing predicted and measured travel times \cite{rayls}. The accuracy of these predictions also depends on how the SSP is represented, motivating travel-time-weighted SSP simplification for ray calculations \cite{sspsimplify}. These methods account for refraction using sound-speed information representative of the propagation environment.

When the current SSP is uncertain, one approach is to estimate the target position and sound-speed parameters jointly. Under an isogradient model, the SSP can be described by a small set of parameters and estimated together with the target position from TOA observations \cite{unified,cheniso}. Robust formulations also address localization under stratification and measurement or model perturbations \cite{ctlsl}. Joint estimation depends on how well the assumed profile family describes the actual environment, while robust formulations depend on the perturbations included in the model. These dependencies become important when a representative SSP is unavailable at localization time.

The accuracy of propagation-model-based localization also depends on reference-station geometry and coordinate accuracy. Information-theoretic analyses and sensor-placement studies characterize how geometry affects positioning accuracy, including in stratified environments \cite{limits,placement,geometry}. Coordinate errors introduce a separate source of mismatch because the assumed propagation endpoints differ from their true positions. This has motivated sensor self-localization with uncertain beacon positions and calibration using imperfectly located emitters \cite{beacon,calibration}. Together, these studies identify two distinct inputs to explicit travel-time prediction: the SSP and the reference-station coordinates. The following subsection examines how data-driven methods handle environmental variability when the current SSP is unavailable during localization.

\subsection{Data-Driven Localization under Environmental Mismatch}
\label{sec:2.2}

Matched-field processing estimates source position by comparing the observed acoustic field with replicas generated for candidate source positions \cite{mfp}. Data-driven methods learn a mapping from acoustic observations to source range or position using training samples \cite{bianco}. For example, covariance features derived from array pressure measurements have been used for source-range classification \cite{niu}, and generalized regression neural networks have been applied to continuous source-position estimation \cite{grnn}. These studies show how acoustic-field features can be used to learn a localization mapping.

Propagation models can generate training samples across different source positions and environmental conditions. A practical challenge is that the simulated environment may differ from the deployment environment. Transfer learning has been used to adapt models trained on synthetic data to deep-sea ranging experiments \cite{wangtransfer}. Simulations have also been used to assess the effects of SSP and seabed-depth mismatch on learned localization \cite{chencnn} and to improve tolerance to environmental mismatch by including a wider range of conditions in training \cite{diniz}. These studies show why the environmental conditions represented in training matter when a learned estimator is applied to a new setting.

Environmental variability can also be addressed through propagation compensation or adaptation during deployment. For underwater sensor networks, broad learning has been combined with sound-speed stratification compensation to account for refraction in the localization process \cite{broad}. Joint source--environment adaptation uses deployment observations to account for discrepancies between the training and test environments \cite{adapt}. In this work, ray tracing under multiple SSPs provides the training observations, and the learned model is evaluated on held-out SSPs with its parameters fixed. This setting assesses localization accuracy when the current SSP is unavailable at inference.

The observations considered here are scalar TOA measurements collected at spatially distributed reference stations. For a given target, each travel-time observation depends on the station location and the propagation environment. The relative positions of two stations provide spatial context for interpreting their paired observations. The resulting design question is how to fuse individual TOA observations with station-pair timing and geometry, which motivates the graph-based methods discussed next.

\subsection{Graph Neural Networks and Geometry--Time Relation Modeling}
\label{sec:2.3}

A graph represents distributed TOA observations by assigning reference stations to nodes and station-pair relationships to edges \cite{bronstein,wu}. Graph convolutional networks use normalized adjacency to aggregate neighboring features \cite{filters,gcn}. Graph attention networks (GATs) make this aggregation feature-dependent by learning weights from node representations \cite{gat}. In the original GAT formulation, the receiving node can affect the attention weights, but it cannot change the ranking of a fixed set of neighbors. GATv2 changes the order of the projection and scoring operations so that the ranking can depend on the receiving node \cite{gatv2}.

Dynamic attention determines how to weight incoming messages. A related design question is how to use information associated with the edges themselves. Relation-aware graph attention incorporates such information into neighborhood aggregation, as illustrated by methods that distinguish semantic relation types in knowledge graphs \cite{attsurvey,multirel}. Edge information can enter the attention score, where it affects the weight assigned to a message, or the message itself, where it adds information to the receiving node's update. These two roles provide a basis for deciding how to use station-pair information in multistation localization.

For a given underwater target, the relative displacement and distance between two reference stations provide spatial context for interpreting their paired TOA observations. Each individual TOA also provides a travel-time observation for the corresponding target--station link. A key design challenge is to use station-pair geometry--time relations to inform both attention weights and message content while giving successive graph layers direct access to individual TOA features.

\section{UGAN for Multistation Underwater TOA Localization}
\label{sec:3}

\subsection{Problem Formulation and TOA Observation Model}
\label{sec:toa_observation_model}

To address three-dimensional underwater localization when the SSP of the current environment is unavailable at inference, we propose UGAN to fuse time-of-arrival (TOA) observations from multiple reference stations with their spatial coordinates. We first define the localization task and the observation model used for the subsequent accuracy analysis.

Consider $N_{\mathrm{tar}}$ underwater targets and $M$ reference stations. Let $\boldsymbol{p}_n\in\mathbb{R}^{3}$ denote the position of target $n$. We assume that the transmission time $t_n^{\mathrm{tx}}$ of each target is known, the transmitter and receiver clocks are synchronized, and the measured arrivals are associated with their respective targets. If $t_{n,i}^{\mathrm{arr}}$ is the measured arrival time at station $i$ for target $n$, the propagation-time observation is
\begin{equation}
	t_{n,i}=t_{n,i}^{\mathrm{arr}}-t_n^{\mathrm{tx}}. \label{eq1}
\end{equation}
Hereafter, $t_{n,i}$ denotes the measured propagation time obtained from the synchronized arrival and transmission timestamps. We refer to it as the TOA observation for brevity.

Let $c(d)$ denote the SSP as a function of depth $d$, and let $\boldsymbol{b}_{n,i}^{\mathrm{true}}$ denote the true position of reference station $i$ for the observation of target $n$. The TOA observation is modeled as
\begin{equation}
	t_{n,i}
	= \tau_{n,i}\!\left(\boldsymbol{p}_n;
	\boldsymbol{b}_{n,i}^{\mathrm{true}},c\right)
	+ \epsilon_{n,i},
\end{equation}
where $\tau_{n,i}$ is the noise-free travel time along the direct acoustic path in the stratified medium and $\epsilon_{n,i}$ is the TOA measurement error. The travel time depends on the target and station positions and on the SSP.

For a given SSP and true station geometry, the errors are modeled as independent zero-mean Gaussian variables across the $M$ links of a target. The variance of each link can depend on target position because the propagation path changes with $\boldsymbol{p}_n$. With the SSP and true station coordinates implicit in the notation, the conditional observation model is
\begin{equation}
	\boldsymbol{t}_n \mid \boldsymbol{p}_n
	\sim \mathcal{N}\!\left(
	\boldsymbol{\tau}_n(\boldsymbol{p}_n),
	\boldsymbol{R}_n(\boldsymbol{p}_n)
	\right),
\end{equation}
where $\boldsymbol{t}_n=[t_{n,1},\ldots,t_{n,M}]^{\mathrm{T}}$,
$\boldsymbol{\tau}_n(\boldsymbol{p}_n)$ stacks the noise-free travel times, and
\begin{equation}
	\boldsymbol{R}_n(\boldsymbol{p}_n)
	= \operatorname{diag}\!\left(
	\sigma_{n,1}^{2}(\boldsymbol{p}_n),\ldots,
	\sigma_{n,M}^{2}(\boldsymbol{p}_n)
	\right).
\end{equation}
This conditional model is used to derive the Cramér--Rao lower bound in the following subsection.

In the localization task, let $\boldsymbol{b}_{n,i}$ denote the coordinates of station $i$ supplied to the estimator. These coordinates may differ from $\boldsymbol{b}_{n,i}^{\mathrm{true}}$ when installation offsets are present. For each target $n$, UGAN estimates $\boldsymbol{p}_n$ from the set of multistation observations
$\{(\boldsymbol{b}_{n,i},t_{n,i})\}_{i=1}^{M}$.

\subsection{Conditional Cramér--Rao Lower Bound}
\label{sec:conditional_crlb}
To provide a theoretical reference for three-dimensional TOA localization, we derive the Cramér--Rao lower bound (CRLB) conditioned on the true SSP, the true reference-station coordinates, and the known noise coefficient. The transmission time is assumed known under the synchronized observation model in Section~\ref{sec:toa_observation_model}. We consider one target and omit the target index for clarity.

Let $\boldsymbol{p}=(x,y,z)^{\mathrm{T}}$ denote the target position. Under the conditional observation model, the TOA vector has mean $\boldsymbol{\tau}(\boldsymbol{p})$ and diagonal covariance $\boldsymbol{R}(\boldsymbol{p})$, with $\sigma_i^2(\boldsymbol{p})$ denoting the variance of link $i$. We define the Jacobian of the mean travel times and the matrix of log-variance gradients as
\begin{equation}
	\label{eq:crlb-jacobians}
	\begin{aligned}
		\boldsymbol{J}(\boldsymbol{p})
		&= \frac{\partial\boldsymbol{\tau}(\boldsymbol{p})}
		{\partial\boldsymbol{p}^{\mathrm{T}}},\\
		[\boldsymbol{G}(\boldsymbol{p})]_{i,:}
		&= \nabla_{\boldsymbol{p}}^{\mathrm{T}}
		\log \sigma_i^2(\boldsymbol{p}),
		\qquad i=1,\ldots,M.
	\end{aligned}
\end{equation}
Both matrices have dimensions $M\times 3$. The $i$th row of $\boldsymbol{J}(\boldsymbol{p})$ describes the sensitivity of the mean travel time on link $i$ to changes in the target coordinates.

For independent Gaussian TOA observations, the Fisher information matrix (FIM) is
\begin{equation}
	\label{eq:crlb-fim}
	\boldsymbol{F}(\boldsymbol{p})
	= \boldsymbol{J}^{\mathrm{T}}(\boldsymbol{p})
	\boldsymbol{R}^{-1}(\boldsymbol{p})\boldsymbol{J}(\boldsymbol{p})
	+ \frac{1}{2}\boldsymbol{G}^{\mathrm{T}}(\boldsymbol{p})
	\boldsymbol{G}(\boldsymbol{p}).
\end{equation}
The first term weights the travel-time sensitivities by the inverse link variances. The second term accounts for information carried by the position dependence of those variances. It vanishes when the variances are locally constant. We retain both terms because the TOA noise model used in the simulations allows the link variances to vary with target position.

Under standard regularity conditions and for a positive-definite $\boldsymbol{F}(\boldsymbol{p})$, any locally unbiased position estimator $\hat{\boldsymbol{p}}$ satisfies
\begin{equation}
	\label{eq:crlb-covariance}
	\operatorname{Cov}(\hat{\boldsymbol{p}}\mid\boldsymbol{p})
	\succeq \boldsymbol{F}^{-1}(\boldsymbol{p}),
\end{equation}
where $\succeq$ denotes the positive semidefinite ordering. Taking the trace gives the bound on the three-dimensional mean squared position error at $\boldsymbol{p}$:
\begin{equation}
	\label{eq:crlb-trace}
	\mathbb{E}\!\left[
	\left\lVert\hat{\boldsymbol{p}}-\boldsymbol{p}\right\rVert_2^2
	\,\middle|\,\boldsymbol{p}
	\right]
	\geq \operatorname{tr}\!\left(\boldsymbol{F}^{-1}(\boldsymbol{p})\right).
\end{equation}
In the simulations, we evaluate this bound at each true target position using the corresponding SSP and true reference-station coordinates. The resulting values provide a conditional accuracy reference for the localization results.

\subsection{UGAN Architecture and Graph Construction}
\label{sec:ugan_architecture}

\begin{figure*}[!t]
	\centering
	\includegraphics[width=\linewidth]{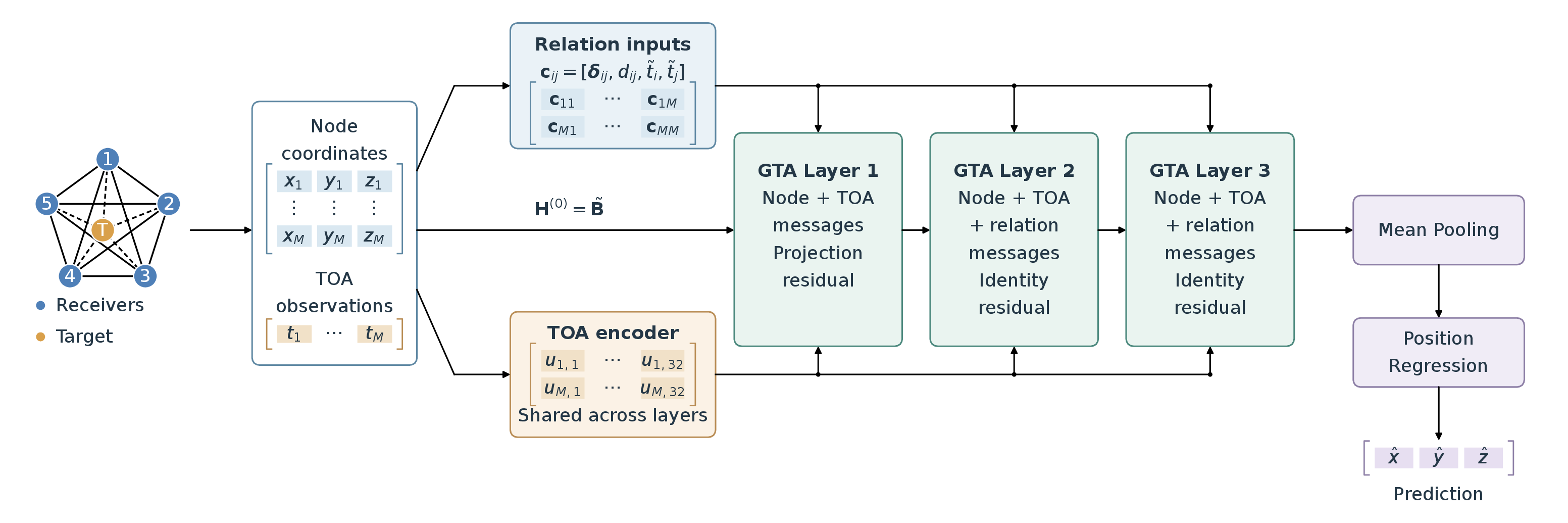}
	\caption{UGAN architecture for 3-D TOA localization with geometry–time relation scoring in all three GTA layers and relation-enhanced messages in layers 2 and 3.}
	\label{fig:architecture}
\end{figure*}

To estimate three-dimensional target positions when the current SSP is unavailable at inference, UGAN organizes the multistation TOA observations and supplied reference-station coordinates defined in Section~\ref{sec:toa_observation_model} into target-specific graphs. Fig.~\ref{fig:architecture} shows how UGAN uses station-pair relations to determine both the weights and the content of aggregated messages. The same difference in arrival times can constrain the target position differently depending on the relative positions of the two stations. Each GTA layer therefore encodes the interstation displacement and distance together with the ordered TOA pair. The resulting geometry–time relations enter attention scoring in all three layers and also contribute to the messages aggregated in the second and third layers. The shared individual TOA encoder supplies each layer with a direct representation of each station’s timing observation.

For target $n$, we construct a fully connected directed graph $\mathcal{G}_n$ with self-loops. Each node represents one of the $M$ reference stations, and the TOA observation at that station is associated with target $n$. An edge $j\to i$ allows receiving node $i$ to aggregate a message from source node $j$. The self-loop includes the station's own message in this aggregation. We omit the target index in the following definitions.

Let $\boldsymbol{b}_i$ denote the coordinates of reference station $i$ supplied to the estimator, and let $L=1000$~m be the spatial scaling factor. The normalized station coordinates $\tilde{\boldsymbol{b}}_i$ initialize the node state:
\begin{equation}
	\boldsymbol{h}_i^{(0)} = \tilde{\boldsymbol{b}}_i = \frac{\boldsymbol{b}_i}{L}. \label{eq09}
\end{equation}

In Fig.~\ref{fig:architecture}, $\tilde{\boldsymbol{B}}$ stacks these normalized station coordinates, so $\boldsymbol{H}^{(0)}=\tilde{\boldsymbol{B}}$.

For the directed edge $j \to i$, we express the location of source node $j$ relative to receiving node $i$ as
\begin{equation}
	\boldsymbol{\delta}_{ij} = \tilde{\boldsymbol{b}}_j - \tilde{\boldsymbol{b}}_i. \label{eq10}
\end{equation}
The corresponding normalized interstation distance is
\begin{equation}
	d_{ij} = \lVert \boldsymbol{\delta}_{ij} \rVert_2. \label{eq11}
\end{equation}
We concatenate the three displacement components and the distance to obtain the four-dimensional geometric edge feature
\begin{equation}
	\boldsymbol{e}_{ij} = [\boldsymbol{\delta}_{ij};\,d_{ij}]. \label{eq12}
\end{equation}
For a self-loop, $\boldsymbol{e}_{ii}=\boldsymbol{0}$. The same geometric edge features are supplied to all three GTA layers, where they are combined with the paired TOA observations defined in the following subsection.

UGAN derives individual TOA features and ordered TOA pairs from the observations for each target. The individual features enter the node messages in all three GTA layers. The ordered TOA pairs are encoded together with the geometric edge features to form station-pair relations. These relations guide attention in all three layers and contribute to message content in the second and third layers. After the final layer, graph-level pooling and regression yield one three-dimensional position estimate per graph. The same trained model processes multiple target-specific graphs in a batch and returns one estimate for each target.

\subsection{Individual and Paired TOA Representations}
For a given target, UGAN uses the same set of TOA observations to construct individual station features and ordered station-pair inputs. We first standardize the TOA observation at station $i$ as
\begin{equation}
	\tilde{t}_i = \frac{t_i-\mu_t}{\sigma_t+\varepsilon}, \label{eq13}
\end{equation}
where $\mu_t$ and $\sigma_t$ are the mean and standard deviation computed over all TOA observations in the training set, and $\varepsilon=10^{-9}$. The same statistics are used for validation and testing.

\begin{figure}[!t]
	\centering
	\includegraphics[width=0.98\columnwidth]{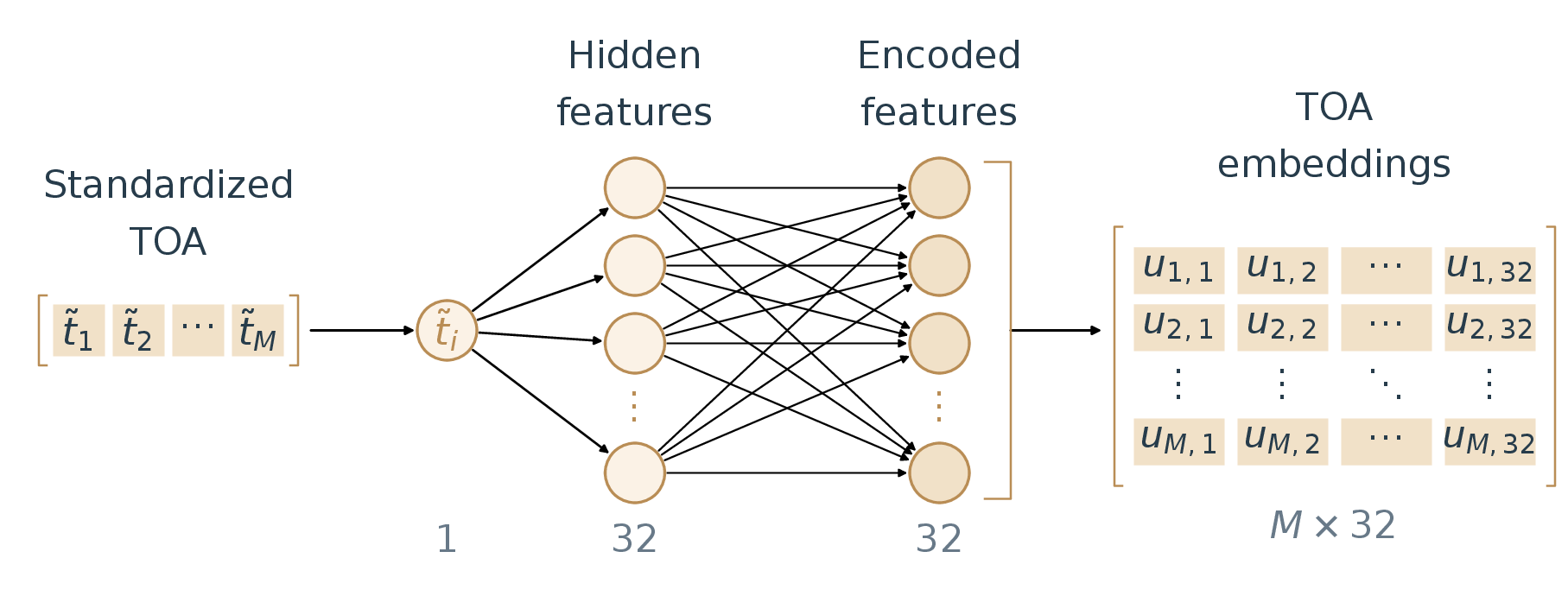}
	\caption{Nonlinear TOA encoder.}
	\label{fig:toa_encoder}
\end{figure}

To retain the timing information of each target--station link, we encode $\tilde{t}_i$ using a nonlinear encoder shared across reference stations. As shown in Fig.~\ref{fig:toa_encoder}, the encoder consists of two bias-free linear transformations, each followed by a GELU activation.
The individual TOA feature is given by
\begin{equation}
	\boldsymbol{u}_i =
	\phi\!\left(\boldsymbol{W}_{\mathrm{t2}}
	\phi\!\left(\boldsymbol{W}_{\mathrm{t1}}\tilde{t}_i\right)\right), \label{eq14}
\end{equation}
where $\phi(\cdot)$ denotes GELU and $\boldsymbol{W}_{\mathrm{t1}}$ and $\boldsymbol{W}_{\mathrm{t2}}$ are trainable weight matrices. The resulting feature $\boldsymbol{u}_i$ is reused by all three GTA layers.

For each directed edge $j\to i$, we form an ordered TOA pair as
\begin{equation}
	\boldsymbol{\tau}_{ij} =
	[\tilde{t}_i;\,\tilde{t}_j],\label{eq15}
\end{equation}
where the observation at receiving node $i$ appears first and that at source node $j$ appears second, following the edge convention in the preceding subsection. The pair retains both standardized TOA values. In the following subsection, each GTA layer encodes this pair using its own parameters and combines it with the corresponding geometric edge feature $\boldsymbol{e}_{ij}$ to obtain a station-pair relation.

\subsection{Geometry--Time Graph Attention}
\label{sec:3.4}
\begin{figure}[!t]
	\centering
	\includegraphics[width=0.98\columnwidth]{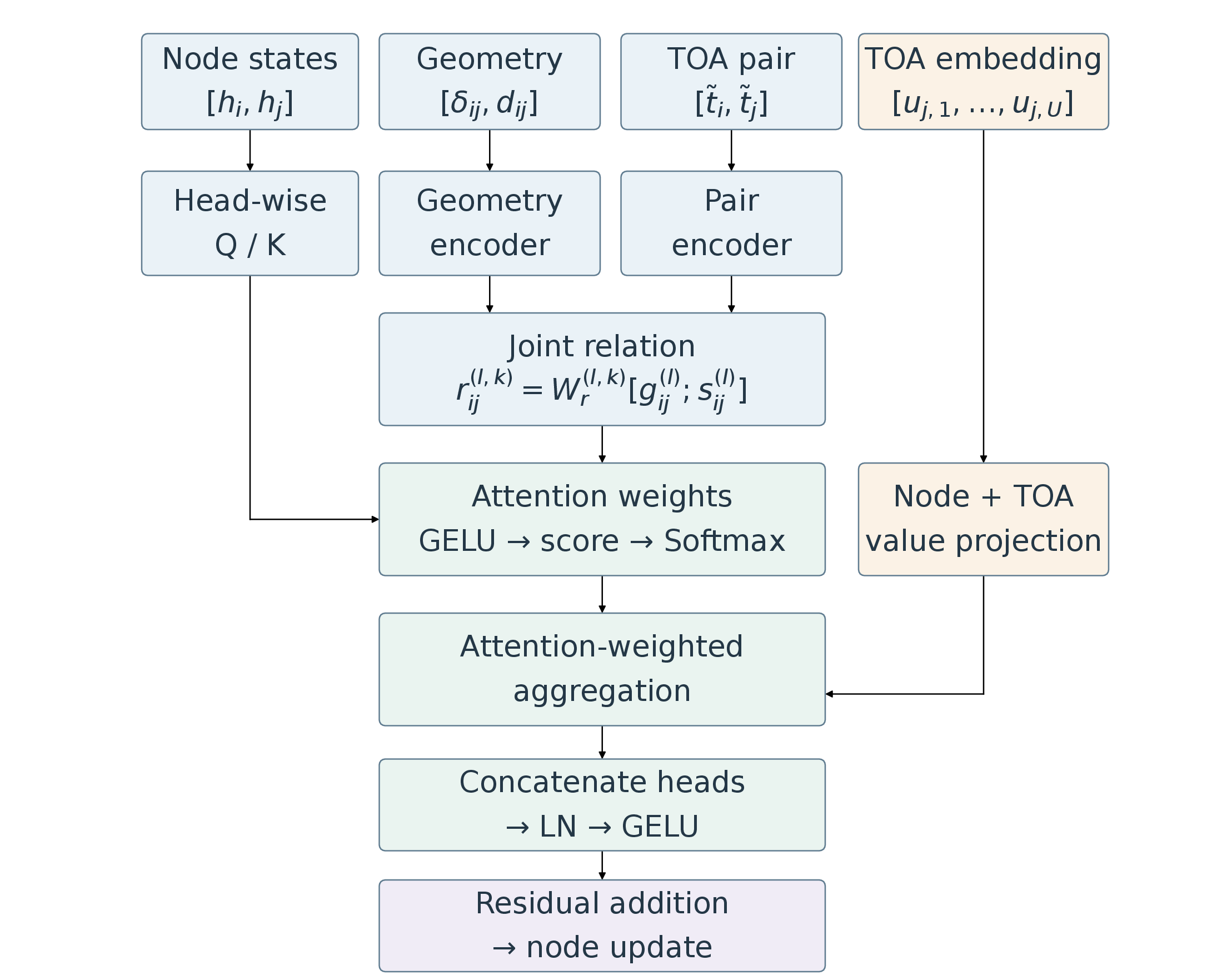}
	\caption{Geometry--time graph attention layer. The residual uses an affine projection in layer 1 and an identity mapping in layers 2--3. Relation messages use $\eta_l\boldsymbol{B}^{(l,k)}$, with $\eta_1=0$ and $\eta_2=\eta_3=1$; self-loop relation features are masked to zero.}
	\label{fig:graph_attention}
\end{figure}

To fuse multistation observations in their geometric context, we design a Geometry--Time Attention (GTA) layer that uses station-pair relations to guide attention and enrich message content. Each layer also has direct access to the individual TOA features defined in the preceding subsection. Fig.~\ref{fig:graph_attention} illustrates the layer architecture. Let $l\in\{1,2,3\}$ index the GTA layers and $k\in\{1,\ldots,K\}$ index the attention heads.

\subsubsection{Station-Pair Relation Encoding}

For the directed edge $j\to i$, layer $l$ separately encodes the geometric edge feature $\boldsymbol{e}_{ij}$ and the ordered TOA pair $\boldsymbol{\tau}_{ij}$ as
\begin{equation}
	\label{eq16}
	{\boldsymbol{g}}_{ij}^{(l)}=\phi({\boldsymbol{W}}_{\mathrm{g2}}^{(l)}\phi({\boldsymbol{W}}_{\mathrm{g1}}^{(l)}{\boldsymbol{e}}_{ij})),
\end{equation}
\begin{equation}
	\label{eq17}
	{\boldsymbol{s}}_{ij}^{(l)}=\phi({\boldsymbol{W}}_{\mathrm{s2}}^{(l)}\phi({\boldsymbol{W}}_{\mathrm{s1}}^{(l)}{\boldsymbol{\tau}}_{ij})),
\end{equation}
where $\phi(\cdot)$ denotes GELU. The two encoders are shared across heads within each layer and have separate parameters across layers. Each head projects the concatenated encodings into its scoring space to obtain an $R$-dimensional relation feature
\begin{equation}
	\label{eq18}
	{\boldsymbol{r}}_{ij}^{(l,k)}=\mathbb{I}_{i\ne j}{\boldsymbol{W}}_{\mathrm{r}}^{(l,k)}[{\boldsymbol{g}}_{ij}^{(l)};{\boldsymbol{s}}_{ij}^{(l)}],
\end{equation}
where $\boldsymbol{W}_{\mathrm{r}}^{(l,k)}$ is a bias-free projection. The indicator $\mathbb{I}_{i\ne j}$ equals one for distinct stations and zero for a self-loop, setting the self-loop relation feature to zero. The self-loop still contributes through the node-state and individual TOA terms.

\subsubsection{Geometry--Time Attention Scoring}
Following the dynamic scoring principle of GATv2 \cite{gatv2}, GTA combines the receiving- and source-node projections before applying the scoring nonlinearity. We obtain the query and key vectors as
% Source equation (3-11)
\begin{equation}
	\label{eq19}
	{\boldsymbol{q}}_{i}^{(l,k)}={\boldsymbol{W}}_{\mathrm{q}}^{(l,k)}{\boldsymbol{h}}_{i}^{(l-1)},
\end{equation}
% Source equation (3-12)
\begin{equation}
	\label{eq20}
	{\boldsymbol{k}}_{j}^{(l,k)}={\boldsymbol{W}}_{\mathrm{k}}^{(l,k)}{\boldsymbol{h}}_{j}^{(l-1)}.
\end{equation}
The station-pair relation is added in the same scoring space. The unnormalized attention score is
\begin{equation}
	\label{eq21}
	{z}_{ij}^{(l,k)}={({\boldsymbol{a}}^{(l,k)})}^{\mathrm{T}}\phi({\boldsymbol{q}}_{i}^{(l,k)}+{\boldsymbol{k}}_{j}^{(l,k)}+{\boldsymbol{r}}_{ij}^{(l,k)}),
\end{equation}
where $\boldsymbol{a}^{(l,k)}$ is a learnable scoring vector. For each receiving node $i$, the scores are normalized over all source nodes, including $i$ itself,
% Source equation (3-14)
\begin{equation}
	\label{eq22}
	{\alpha}_{ij}^{(l,k)}=\frac{\operatorname{exp}({z}_{ij}^{(l,k)})}{\sum_{r=1}^{M}\operatorname{exp}({z}_{ir}^{(l,k)})}.
\end{equation}
The first layer computes attention from coordinate-based node states and station-pair geometry--time relations. Later layers recompute the weights using node states updated by the preceding layer.

\subsubsection{Message Construction and Aggregation}

Each head constructs a node message from the source-node state and its individual TOA feature $\boldsymbol{u}_j$. The two inputs are projected separately and added as
% Source equation (3-15)
\begin{equation}
	\label{eq23}
	{\boldsymbol{v}}_{j}^{(l,k)}={\boldsymbol{W}}_{\mathrm{v}}^{(l,k)}{\boldsymbol{h}}_{j}^{(l-1)}+{\boldsymbol{b}}_{\mathrm{v}}^{(l,k)}+{\boldsymbol{W}}_{\mathrm{u}}^{(l,k)}{\boldsymbol{u}}_{j}.
\end{equation}
The projection $\boldsymbol{W}_{\mathrm{u}}^{(l,k)}$ gives each head in every layer direct access to the individual TOA feature. The head output at node $i$ is
% Source equation (3-16)
\begin{equation}
	\label{eq24}
	\begin{aligned}{\boldsymbol{o}}_{i}^{(l,k)}=\sum_{j=1}^{M}{\alpha}_{ij}^{(l,k)}{\boldsymbol{v}}_{j}^{(l,k)} {}+{\eta}_{l}{\boldsymbol{B}}^{(l,k)}\sum_{j=1}^{M}{\alpha}_{ij}^{(l,k)}{\boldsymbol{r}}_{ij}^{(l,k)},
	\end{aligned}
\end{equation}
where $\eta_1=0$ and $\eta_2=\eta_3=1$. The trainable projection $\boldsymbol{B}^{(l,k)}$ maps the aggregated relation feature into the node-message space and is initialized to zero. Thus, station-pair relations guide attention in all three layers and contribute directly to message content in the second and third layers.

\subsubsection{Multihead Fusion and Node Update}

The outputs of the $K$ attention heads are concatenated and combined with a learnable bias $\boldsymbol{\beta}^{(l)}$,
% Source equation (3-17)
\begin{equation}
	\label{eq25}
	{\boldsymbol{m}}_{i}^{(l)}={\operatorname{Concat}}_{k=1}^{K}({\boldsymbol{o}}_{i}^{(l,k)})+{\boldsymbol{\beta}}^{(l)}.
\end{equation}
The layer applies layer normalization followed by GELU before adding the residual contribution,
% Source equation (3-18)
\begin{equation}
	\label{eq26}
	{\boldsymbol{h}}_{i}^{(l)}=\phi(\operatorname{LayerNorm}({\boldsymbol{m}}_{i}^{(l)}))+{P}_{l}({\boldsymbol{h}}_{i}^{(l-1)}),
\end{equation}
where $P_1$ is an affine projection that matches the initial coordinate features to the output dimension, while $P_2$ and $P_3$ are identity mappings. The final node states are passed to the graph-level readout for position regression.

\begin{remark}[Individual and Paired TOA Representations]
	\label{rem:toa-representations}
	The individual and paired representations are derived from the same TOA observations. The individual branch supplies each GTA layer with a direct representation of each target--station travel time. The paired branch places two TOA observations in their	interstation geometric context and contributes to attention scoring	in all layers and to message content in layers 2 and 3. The paired branch remains active in the TOA-branch ablation, so this ablation measures the incremental benefit of the direct individual path.
\end{remark}

\subsection{Graph-Level Readout, Position Regression, and Training}
\label{sec:3.5}
Each observation graph contains the multistation measurements associated with one target. After the third GTA layer, we apply global mean pooling to the final node states to obtain a graph-level representation
% Source equation (3-19)
\begin{equation}
	\label{eq27}
	\boldsymbol{g}=\frac{1}{M}\sum_{i=1}^{M}{\boldsymbol{h}}_{i}^{(3)},
\end{equation}
where $\boldsymbol{h}_i^{(3)}$ is the final feature vector of reference station $i$. Mean pooling produces a fixed-dimensional representation that is invariant to the ordering of the station features.

A position regressor $f_{\mathrm{reg}}$ maps $\boldsymbol{g}$ to a normalized three-dimensional position estimate
% Source equation (3-20)
\begin{equation}
	\label{eq28}
	{\hat{\boldsymbol{p}}}_{\mathrm{norm}}={f}_{\mathrm{reg}}(\boldsymbol{g}).
\end{equation}
The regressor has two hidden layers and a linear output layer. The first hidden layer uses layer normalization followed by GELU, and the second uses GELU. We recover the position estimate in meters using the spatial scaling factor $L$ defined in the graph construction:
% Source equation (3-21)
\begin{equation}
	\label{eq29}
	\hat{\boldsymbol{p}}=L{\hat{\boldsymbol{p}}}_{\mathrm{norm}}.
\end{equation}
In batched inference, the readout and regression are applied to each target-specific graph, yielding one three-dimensional position estimate per graph.

\begin{proposition}[Invariance to Reference-Station Ordering]
	\label{prop:station-permutation}
	For a given target, let $f_{\mathrm{UGAN}}$ denote the complete inference mapping from the observation matrix $\mathbf{X}\in\mathbb{R}^{M\times 4}$ to the estimated position 	$\hat{\mathbf{p}}\in\mathbb{R}^{3}$, where $\mathbf{X}_{i,:}=[\mathbf{b}_i^{\mathsf T},\,t_i]$.	For any permutation $\pi$ of the reference stations, let $\mathbf{P}_{\pi}$ be the corresponding permutation matrix, so that $(\mathbf{P}_{\pi}\mathbf{X})_{\pi(i),:}=\mathbf{X}_{i,:}$. Then
	\begin{equation}
		f_{\mathrm{UGAN}}(\mathbf{P}_{\pi}\mathbf{X})
		= f_{\mathrm{UGAN}}(\mathbf{X}).
		\label{eq:ugan-permutation-invariance}
	\end{equation}
\end{proposition}

\begin{IEEEproof}
	Let $\mathbf{X}_{\pi}=\mathbf{P}_{\pi}\mathbf{X}$, and use the subscript $\pi$ for quantities computed from $\mathbf{X}_{\pi}$.	The coordinate normalization and shared TOA encoder give
	$\mathbf{H}_{\pi}^{(0)}=\mathbf{P}_{\pi}\mathbf{H}^{(0)}$ and $(\mathbf{u}_{\pi})_{\pi(i)}=\mathbf{u}_i$. For each directed edge $j\to i$, the geometric features, ordered TOA pair, and
	self-loop mask are reindexed together. Hence, for every layer $l$ and head $k$,
	\begin{equation}
		(\mathbf{r}_{\pi}^{(l,k)})_{\pi(i),\pi(j)}
		= \mathbf{r}_{ij}^{(l,k)}.
	\end{equation}
	
	Suppose that $\mathbf{H}_{\pi}^{(l-1)}=\mathbf{P}_{\pi}\mathbf{H}^{(l-1)}$. For a given head, the shared projections and row-wise softmax over source nodes yield 
	\begin{equation}
		\begin{aligned}
			\mathbf{A}_{\pi}^{(l,k)}
			&= \mathbf{P}_{\pi}\mathbf{A}^{(l,k)}
			\mathbf{P}_{\pi}^{\mathsf T},\\
			\mathbf{V}_{\pi}^{(l,k)}
			&= \mathbf{P}_{\pi}\mathbf{V}^{(l,k)},
		\end{aligned}
	\end{equation}
	where $\mathbf{A}^{(l,k)}$ contains the attention weights and $\mathbf{V}^{(l,k)}$ stacks the node messages. The weighted node messages therefore satisfy $\mathbf{A}_{\pi}^{(l,k)}\mathbf{V}_{\pi}^{(l,k)} =\mathbf{P}_{\pi}\mathbf{A}^{(l,k)}\mathbf{V}^{(l,k)}$. The relation-message sum in (24) is reindexed in the same way, since both $\alpha_{ij}^{(l,k)}$ and $\mathbf{r}_{ij}^{(l,k)}$ are reindexed by $(i,j)\mapsto(\pi(i),\pi(j))$.	The node-wise operations in (25)--(26) preserve this reindexing, so $\mathbf{H}_{\pi}^{(l)}=\mathbf{P}_{\pi}\mathbf{H}^{(l)}$. By induction, this holds through layer $3$.
	
	Finally, using $\mathbf{P}_{\pi}^{\mathsf T}\mathbf{1}_M =\mathbf{1}_M$, mean pooling gives
	\begin{equation}
		\mathbf{g}_{\pi}
		= \frac{1}{M}(\mathbf{H}_{\pi}^{(3)})^{\mathsf T}\mathbf{1}_M
		= \frac{1}{M}(\mathbf{H}^{(3)})^{\mathsf T}\mathbf{1}_M
		= \mathbf{g}.
	\end{equation}
	Applying the same position regressor and spatial scale yields $L f_{\mathrm{reg}}(\mathbf{g}_{\pi}) = L f_{\mathrm{reg}}(\mathbf{g})$, which proves the claim.
\end{IEEEproof}

We train UGAN by minimizing the mean squared Euclidean position error over a mini-batch of $B$ observation graphs:
% Source equation (3-22)
\begin{equation}
	\label{eq30}
	{\mathcal{L}}_{\mathrm{pos}}=\frac{1}{B}\sum_{n=1}^{B}{{\lVert{\hat{\boldsymbol{p}}}_{n}-{\boldsymbol{p}}_{n}\rVert}_{2}}^{2},
\end{equation}
where $\hat{\boldsymbol{p}}_n$ and $\boldsymbol{p}_n$ are the estimated and ground-truth positions, respectively, for the target associated with the $n$th graph in the batch. We select the model checkpoint using the validation localization RMSE.

\subsection{Localization Procedure and Computational Complexity}
\label{sec:3.6}

Algorithm~\ref{alg:ugan-localization} summarizes batched inference with the trained UGAN. The target observations are partitioned into batches of at most $B_{\mathrm{inf}}$ targets, with one observation graph constructed per target. Each batch is processed in one forward pass using the trained model parameters and the TOA standardization statistics computed from the training set. Attention, message aggregation, and pooling operate separately within each graph, yielding one three-dimensional position estimate per target. Individual TOA features are computed once per graph and reused across all three GTA layers. Station-pair relation features are computed separately in each layer using its own parameters. Ray tracing is used offline to generate the simulation data.

\begin{algorithm}[!t]
	\caption{Batched UGAN localization with target-specific graphs}
	\label{alg:ugan-localization}
	\begin{algorithmic}[1]
		\REQUIRE Observation sets $\mathcal{S}_n = 	\{(\mathbf{b}_{n,i},t_{n,i})\}_{i=1}^{M}$ for $n=1,\ldots,N_{\mathrm{tar}}$, with $t_{n,i}$ defined in \eqref{eq1}.
		\REQUIRE Trained UGAN parameters, TOA statistics $\mu_t$ and $\sigma_t$, stabilizer $\varepsilon$, spatial scale $L$, and maximum inference batch size $B_{\mathrm{inf}}$.
		\ENSURE Position estimates $\{\hat{\mathbf{p}}_n\}_{n=1}^{N_{\mathrm{tar}}}$.
		
		\STATE Partition the targets into batches of at most $B_{\mathrm{inf}}$. 
		\FOR{each batch with target index set $\mathcal{I}$}
		\STATE Assemble the station coordinates and TOAs into batched tensors.
		\STATE Initialize node states from scaled coordinates using \eqref{eq09}.
		\STATE Standardize the TOA observations using \eqref{eq13}.
		\STATE Construct one fully connected graph with self-loops per target.
		\STATE Compute geometric edge features using \eqref{eq10}--\eqref{eq12}.
		\STATE Encode individual TOAs using \eqref{eq14} and retain the features
		for all three layers.
		\STATE Form ordered TOA pairs using \eqref{eq15}.
		\FOR{$l=1,2,3$}
		\STATE Encode station-pair relations and mask self-loops
		using \eqref{eq16}--\eqref{eq18}.
		\STATE Compute attention scores and normalize over source nodes
		within each graph using \eqref{eq19}--\eqref{eq22}.
		\STATE Construct node messages using \eqref{eq23}.
		\STATE Aggregate node and relation messages using \eqref{eq24}, with
		$(\eta_1,\eta_2,\eta_3)=(0,1,1)$.
		\STATE Concatenate heads and update node states using \eqref{eq25}--\eqref{eq26}.
		\ENDFOR
		\STATE Mean-pool the final node states within each graph using \eqref{eq27}.
		\STATE Regress each normalized target position using \eqref{eq28}.
		\STATE Restore each estimated position in meters using \eqref{eq29}.
		\ENDFOR
		\RETURN $\{\hat{\mathbf{p}}_n\}_{n=1}^{N_{\mathrm{tar}}}$
	\end{algorithmic}
\end{algorithm}

We analyze the arithmetic cost of one forward pass on a single graph. Let $H=3$ be the number of GTA layers and $E=M^2$ the number of directed edges including self-loops. Layer $l$ has input node dimension $D_{l-1}$, $K$ heads, and message dimension $d_l$ per head, so its output dimension is $D_l=Kd_l$. Let $R$ be the per-head scoring dimension and $U$ the dimension of the individual TOA feature. Denote the operation count of the shared TOA encoder for one station by $C_{\mathrm{toa}}$. For layer $l$, let $Q_l$ be the combined cost of the geometric and paired-TOA encoders for one edge, and let $S_l$ be the dimension of their concatenated output. The two encoders operate once per edge and layer, rather than once per head. For a dense mapping from $a$ to $b$ features, the leading arithmetic cost is $O(ab)$; thus, the encoder costs are determined by the sums of the products of adjacent layer widths, with elementwise activation costs included.

The layer cost can be written as
\begin{equation}
\label{eq:gta-computational-cost}
\begin{aligned}
T_l=O\bigl(&MK[D_{l-1}(2R+d_l)+Ud_l]\\
          &+E[Q_l+KS_lR+K(R+d_l)]\\
          &+\eta_l MKRd_l+C_{\mathrm{res},l}\bigr).
\end{aligned}
\end{equation}
The first line accounts for query, key, and value projections and individual TOA injection. The second accounts for edge encoders, head-specific relation projections, attention scoring, and weighted aggregation. The third includes the relation-message projection and residual mapping. Here, $C_{\mathrm{res},1}=O(MD_0D_1)$ for the first-layer affine residual, with $D_0=3$, while $C_{\mathrm{res},l}=O(MD_l)$ for the identity residual additions in the remaining layers. Attention normalization costs $O(EK)$, and node normalization and activation cost $O(MKd_l)$; these are absorbed by the displayed terms. The relation-aggregation contribution is absent in layer one, without changing the stated asymptotic order because relation-based scoring remains active.

In \eqref{eq24}, the weighted relation features are summed before applying $\boldsymbol B^{(l,k)}$. By linearity, this is equivalent to projecting each relation feature before aggregation. It reduces the relation-projection cost from $O(EKRd_l)$ to $O(MKRd_l)$ in layers two and three. Relation aggregation still costs $O(EKR)$, and the edge encoders and attention scores still process all $E$ edges. Similarly, reusing $\boldsymbol u_i$ across layers requires only one evaluation of the individual TOA encoder per station, while retaining the separate injection projections at each layer.

With $C_{\mathrm{reg}}$ denoting the cost of the graph-level regression network, the total single-graph cost is
\begin{equation}
\label{eq:ugan-computational-cost}
T_{\mathrm{graph}}=O(M^2+MC_{\mathrm{toa}})
                  +\sum_{l=1}^{H}T_l
                  +O(MD_H+C_{\mathrm{reg}}).
\end{equation}
The first term covers input preprocessing, graph features, and individual TOA encoding; the final term covers mean pooling and regression. For fixed layer counts, head counts, and feature widths, the cost grows as $O(M^2)$ because the graph is fully connected. This scaling does not imply that edge operations dominate the numerical cost in the five-station configuration, where the feature projections also contribute. Shared encoders and graph-level mean pooling make the trainable parameter count independent of $M$ and $N_{\mathrm{tar}}$ for a fixed architecture; this architectural property alone does not establish localization accuracy for unseen station configurations.

For $N_{\mathrm{tar}}$ independent target graphs, the total arithmetic cost is $O(N_{\mathrm{tar}}T_{\mathrm{graph}})$. Batching permits parallel execution but does not imply the same scaling of elapsed time, which also depends on the hardware and implementation. With fixed feature widths, explicitly storing edge relations and attention intermediates requires $O(M^2)$ activation memory per graph, in addition to model parameters; a batch of $B_{\mathrm{inf}}$ graphs requires $O(B_{\mathrm{inf}}M^2)$ such memory, where $B_{\mathrm{inf}}$ is the inference batch size. Section~\ref{sec:computational-cost} reports parameter counts and MACs for the evaluated configurations. Those MAC counts exclude activation, normalization, and softmax operations, whereas the asymptotic analysis above includes their arithmetic orders.

\section{Performance Evaluation and Discussion}
\label{sec:4}
\subsection{Simulation Setup and Evaluation Protocol}
\label{sec:4.1}

\subsubsection{Simulation Region and Reference-Station Deployment}
\begin{figure}[!htbp]
	\centering
	\includegraphics[width=0.76\columnwidth]{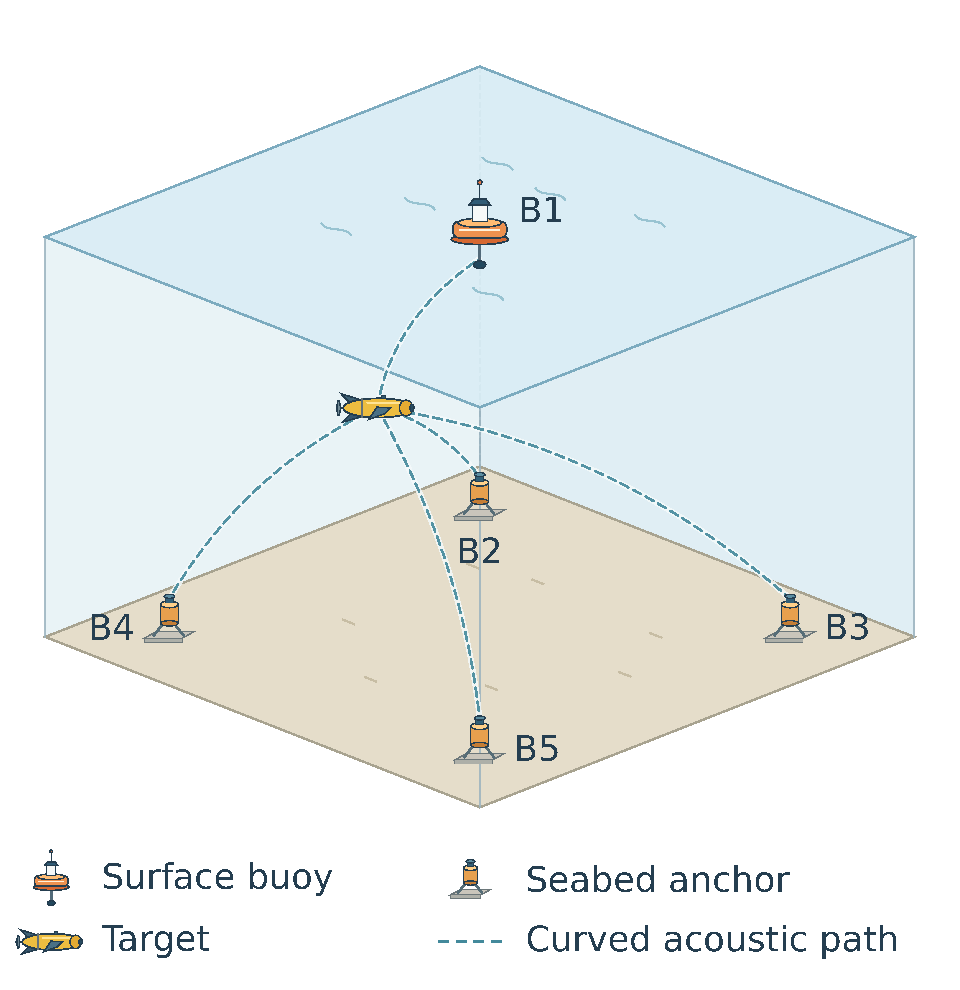}
	\caption{Underwater localization scenario and five-station deployment. The AUV and curved acoustic paths are schematic; hardware icons are not to scale.}
	\label{fig:scenario}
\end{figure}
We consider a cubic localization region with a side length of $1000$~m. As shown in Fig.~\ref{fig:scenario}, the nominal deployment comprises one surface buoy at the center of the upper face and four seabed anchors at the bottom corners. The stations
span two depth levels and provide position constraints from different directions \cite{placement,geometry}.

The horizontal buoy position is sampled uniformly within a disk of radius $10$~m centered at its nominal location. The horizontal installation offset of each seabed anchor is drawn independently from $\mathcal{N}(\boldsymbol{0},\sigma_a^2\boldsymbol{I}_2)$, where $\sigma_a=3$~m and $\boldsymbol{I}_2$ is the $2\times2$ identity matrix. The same realization of the four anchor offsets is used throughout training, validation, and testing. The true station coordinates are used to generate the TOA observations. For UGAN, the supplied coordinates are the nominal seabed-anchor coordinates and the realized surface-buoy coordinates after drift.

\subsubsection{Ray-Traced TOA Generation and Measurement Noise}
\begin{figure}[!htbp]
	\centering
	\subfloat[]{\includegraphics[width=0.65\linewidth]{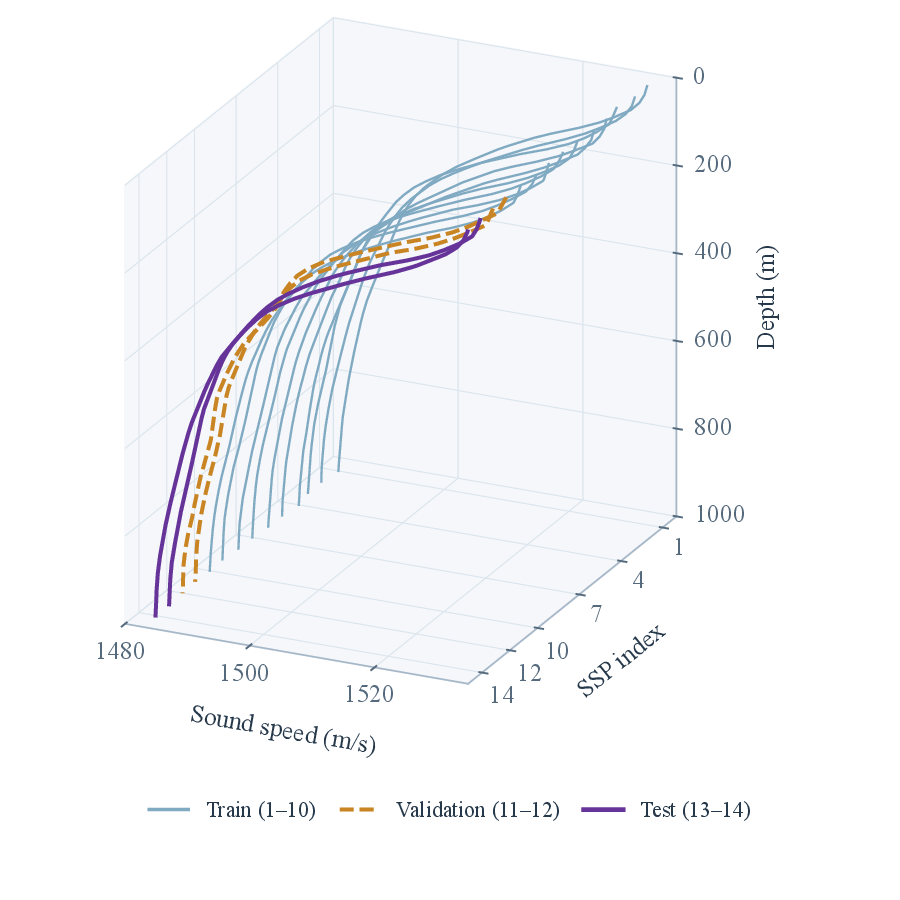}}
	\\
	\subfloat[]{\includegraphics[width=0.65\linewidth]{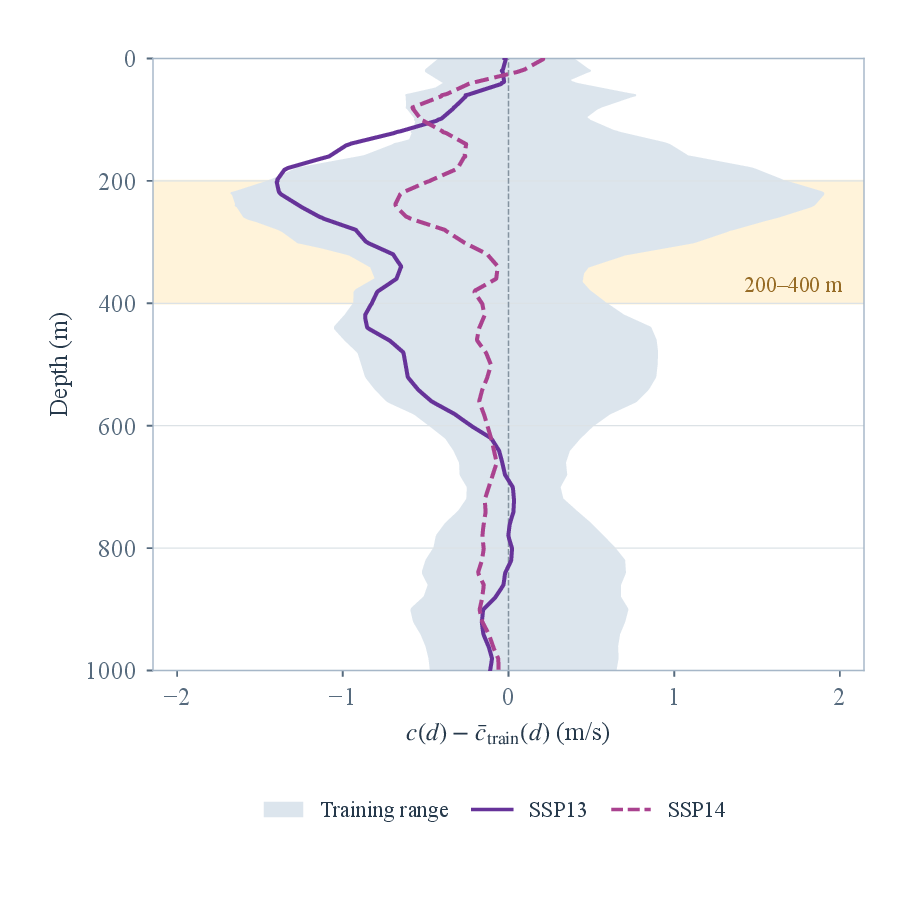}}
	\caption{SSPs used in the ray-tracing simulations. (a) Three-dimensional waterfall plot of the 14 SSPs, arranged by profile index and grouped into training, validation, and test sets. (b) Sound speed deviations of the two test profiles, SSP13 and SSP14, from the training mean at each depth. The gray band spans the minimum and maximum training deviations at each depth, and the pale yellow band highlights the 200–400 m depth interval.}
	\label{fig:ssp}
\end{figure}

Fig.~\ref{fig:ssp}(a) shows the 14 SSPs sampled from the South China Sea and used in the simulations. Fig.~\ref{fig:ssp}(b) shows the deviations of the two test profiles from the training mean as a function of depth, expressed as $c(d)-\bar{c}_{\mathrm{train}}(d)$, where $\bar{c}_{\mathrm{train}}(d)$ is the mean sound speed of the ten training profiles at depth $d$. Near 200~m, SSP13 shows a larger negative deviation than SSP14. For each simulated sample, the selected SSP is held fixed and represented by a continuous piecewise-linear function $c(d)$, where
$d=-z$ denotes depth. We compute the noise-free TOA along a depth-monotone direct ray. Let $q_i$ denote the horizontal slowness of the ray connecting the target and station $i$, and let $R_i$ denote their horizontal separation. For the ray segment within SSP layer $k$, $a_{ik}$ and $b_{ik}$ denote the depths of its shallower and deeper endpoints, respectively. The horizontal slowness satisfies
\begin{equation}
	\sum_k \int_{a_{ik}}^{b_{ik}}
	\frac{q_i c(d)}{\sqrt{1-q_i^2c^2(d)}}\,\mathrm{d}d
	= R_i,
\end{equation}
subject to $0\le q_i c(d)<1$ along the ray. The corresponding
noise-free travel time is
\begin{equation}
	\tau_i =
	\sum_k \int_{a_{ik}}^{b_{ik}}
	\frac{1}{c(d)\sqrt{1-q_i^2c^2(d)}}\,\mathrm{d}d.
\end{equation}

For the noise model, we partition the ray into $1$~m depth intervals. Let $\Delta\tau_{i,m}$ denote the noise-free travel time of link $i$ within interval $m$. Independent zero-mean Gaussian timing perturbations with standard deviations $\gamma\Delta\tau_{i,m}$ are assigned to the intervals and summed along each link. The resulting link-level TOA error $\epsilon_i$ in Section~\ref{sec:toa_observation_model} has variance
\begin{equation}
	\label{eq:crlb-noise-variance}
	\sigma_i^2(\boldsymbol{p})
	= \gamma^2 \sum_m
	\bigl[\Delta\tau_{i,m}(\boldsymbol{p})\bigr]^2,
\end{equation}
where $\gamma$ is a dimensionless noise coefficient. In the simulation, each interval contributes a timing perturbation with standard deviation proportional to its noise-free travel time. Because the interval perturbations are independent, their variances add along the link. An interval with a longer travel time therefore contributes more to the link-level TOA variance, with the contribution proportional to the square of that travel time. We consider $\gamma\in\{0.002,0.005,0.01\}$ and use the same target positions and station geometries across the three noise levels.

\subsubsection{Data Partition}
The data are partitioned by SSP. SSP1--SSP10 are used for training, SSP11--SSP12 for validation, and SSP13--SSP14 for testing. Each sample contains the five TOA observations, the station
coordinates supplied to the estimator, and the ground-truth target position. For UGAN, Proposition~\ref{prop:station-permutation} establishes that the position estimate is invariant to the order of the reference-station records, provided that each TOA observation remains paired with its corresponding station coordinates. At each noise level, the test set contains 4000 samples, with 2000 samples generated under each test SSP. The test SSPs are held out from both training and validation. This protocol evaluates generalization to held-out profiles under the same reference-station deployment and fixed anchor offsets.

\subsubsection{UGAN Training Configuration}
Table~\ref{tab:training-config} summarizes the data sizes and training settings of UGAN. Candidate target positions are sampled independently and uniformly in the simulation cube; samples are retained when valid depth-monotone direct paths exist to all stations. The checkpoint with the lowest validation localization RMSE is used for testing.
\begin{table}[!t]
\caption{Dataset sizes and UGAN training configuration}
\label{tab:training-config}
\centering
\papertableformat
\begin{tabular}{@{}p{0.39\columnwidth}p{0.55\columnwidth}@{}}
\toprule
Item & Setting \\
\midrule
Training / validation samples & 20,000 / 4,000 per noise level \\
Target-position proposals & Uniform in $[0,1000]^2\times[-1000,0]$~m \\
Optimizer & AdamW; weight decay $0$ \\
Initial learning rate & $5\times10^{-4}$ \\
Learning-rate schedule & 10-epoch linear warm-up, then cosine decay to $10^{-7}$ \\
Training batch size & 128 \\
Maximum training epochs & 800 \\
Early stopping & No improvement in validation RMSE for 300 consecutive epochs \\
\bottomrule
\end{tabular}
\end{table}

\subsubsection{Evaluation Protocol and Metrics}
The constant-speed nonlinear least-squares (NLS) baseline converts the TOA observations to ranges using a sound speed of $1500$~m/s and uses the nominal seabed-anchor coordinates and realized buoy coordinates. The learning-based methods and constant-speed NLS are evaluated without access to the test SSP.

The three-dimensional localization error is measured as the Euclidean distance between the estimated and true target positions. The root mean square error (RMSE) over the test set is defined as
\begin{equation}
	\label{eq:test-localization-rmse}
	\mathrm{RMSE}
	= \sqrt{\frac{1}{N}\sum_{n=1}^{N}
		\left\lVert\hat{\boldsymbol{p}}_n-\boldsymbol{p}_n\right\rVert_2^2},
\end{equation}
where $N$ is the number of test samples and $\hat{\boldsymbol{p}}_n$ and $\boldsymbol{p}_n$ denote the estimated and true positions of sample $n$, respectively. The 95th percentile (P95) is computed from the same set of Euclidean position errors.

For each test sample, we evaluate the conditional CRLB derived in Section~\ref{sec:conditional_crlb} using its true target position, SSP, true station coordinates, and noise coefficient.
The test-set summary is
\begin{equation}
	\label{eq:test-aggregate-crlb}
	b_{\mathrm{CRLB}}
	= \sqrt{\frac{1}{N}\sum_{n=1}^{N}
		\operatorname{tr}\!\left(\boldsymbol{F}_n^{-1}\right)},
\end{equation}
where $\boldsymbol{F}_n$ is the Fisher information matrix evaluated for sample $n$. This quantity is the square root of the average conditional MSE bound and provides a theoretical reference for the test-set RMSE.

\subsection{Localization Accuracy and Error Analysis}
\label{sec:4.2}
Table~\ref{tab:performance} compares the three-dimensional localization RMSE of constant-speed NLS and the learning-based methods on the held-out test set described in Section~\ref{sec:4.1}. The corresponding test SSP is unavailable to all methods in this comparison. RMSE is calculated using \eqref{eq:test-localization-rmse}. For the learning-based methods, the entries give the mean and sample standard deviation of the test-set RMSE.

\begin{table}[!t]
	\caption{Three-dimensional localization RMSE without test SSP input and a conditional CRLB reference}
	\label{tab:performance}
	\centering
	\papertableformat
	\begin{tabular*}{\columnwidth}{@{\extracolsep{\fill}}lccc@{}}
		\toprule
		Method & $\gamma=0.002$ & $\gamma=0.005$ & $\gamma=0.01$ \\
		& RMSE / m & RMSE / m & RMSE / m \\
		\midrule
		NLS~\cite{beckrls} & 13.810 & 13.819 & 13.846 \\
		MLP & $1.293\pm0.020$ & $1.404\pm0.022$ & $1.749\pm0.017$ \\
		GCN~\cite{gcn} & $1.074\pm0.039$ & $1.218\pm0.019$ & $1.603\pm0.024$ \\
		GAT~\cite{gat} & $0.822\pm0.041$ & $0.987\pm0.041$ & $1.417\pm0.043$ \\
		GATv2~\cite{gatv2} & $0.640\pm0.039$ & $0.836\pm0.037$ & $1.281\pm0.033$ \\
		UGAN & $0.307\pm0.011$ & $0.577\pm0.018$ & $0.989\pm0.018$ \\
		\midrule
		Conditional CRLB & 0.163 & 0.407 & 0.809 \\
		\bottomrule
	\end{tabular*}
\end{table}

\begin{figure}[!htbp]
	\centering
	\setcounter{subfigure}{0}
	\subfloat[$\gamma=0.002$]{\includegraphics[width=0.66\linewidth]{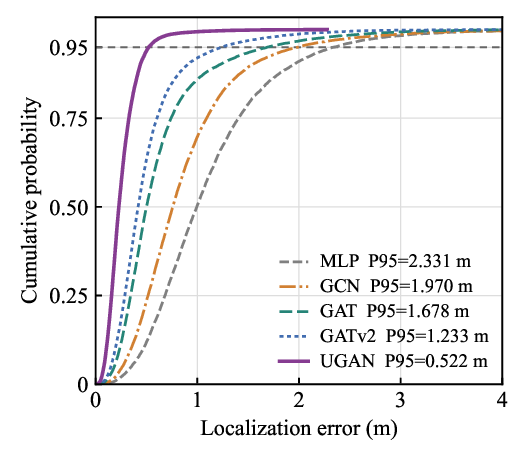}}
	\\
	\subfloat[$\gamma=0.005$]{\includegraphics[width=0.66\linewidth]{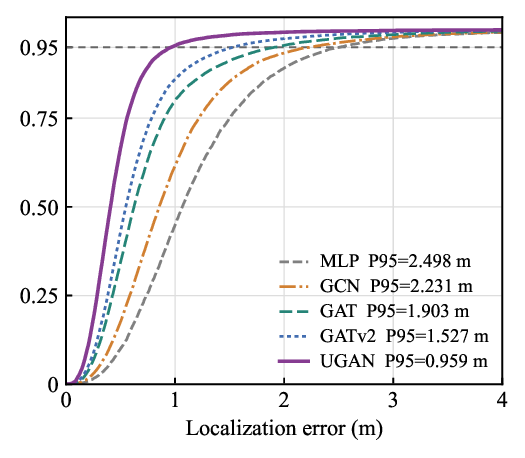}}
	\\
	\subfloat[$\gamma=0.010$]{\includegraphics[width=0.66\linewidth]{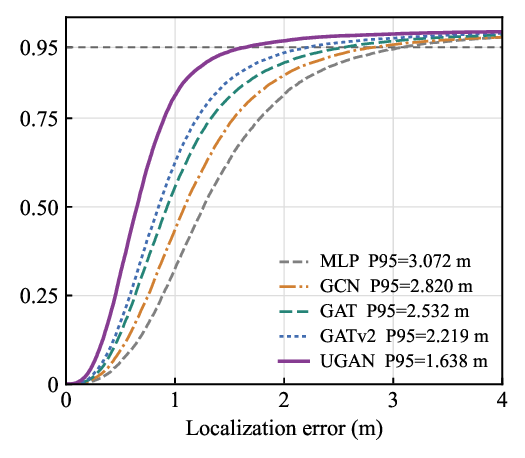}}
	\caption{Mean CDFs and mean 95th percentiles of the three-dimensional position errors for the learning-based methods. Panels (a), (b), and (c) correspond to $\gamma=0.002$, $0.005$, and $0.01$, respectively.}
	\label{fig:cdf_p95}
\end{figure}

The conditional CRLB reference in Table~\ref{tab:performance}, aggregated over the same $4000$ test samples using \eqref{eq:test-aggregate-crlb}, is $0.163$, $0.407$, and $0.809$~m at $\gamma=0.002$, $0.005$, and $0.01$, respectively. UGAN's RMSE remains above these reference values at all three noise levels. The bound is conditioned on the true SSP, true station coordinates, and known noise coefficient and applies to locally unbiased estimators. It therefore provides an ideal-information accuracy reference, without establishing an attainable error limit for the trained network under its available inputs.

UGAN yields the lowest mean RMSE among the evaluated localization methods at all three noise levels, with values of $0.307$, $0.577$, and $0.989$~m, respectively. Relative to GATv2, the learning-based baseline with the lowest RMSE, these values represent reductions of $52.0\%$, $31.0\%$, and $22.8\%$. The relative reduction decreases as the TOA noise level increases. The RMSE of constant-speed NLS varies from $13.810$ to $13.846$~m across the three noise levels. This limited variation is consistent with a substantial contribution from systematic mismatch in the straight-ray propagation model, the
supplied anchor coordinates, or both.

\begin{figure}[!t]
	\setcounter{subfigure}{0}
	\centering
	\subfloat[$\gamma=0.002$]{\includegraphics[width=0.7\linewidth]{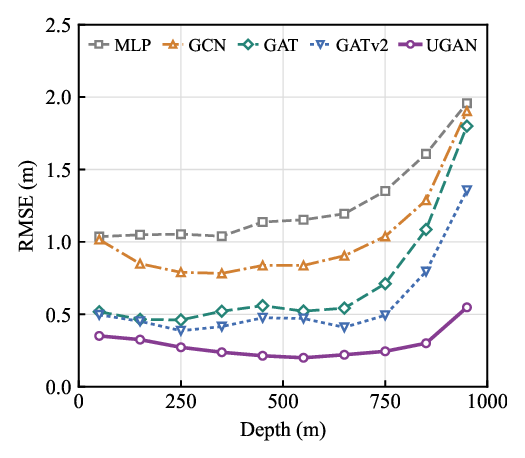}}\\
	\subfloat[$\gamma=0.005$]{\includegraphics[width=0.7\linewidth]{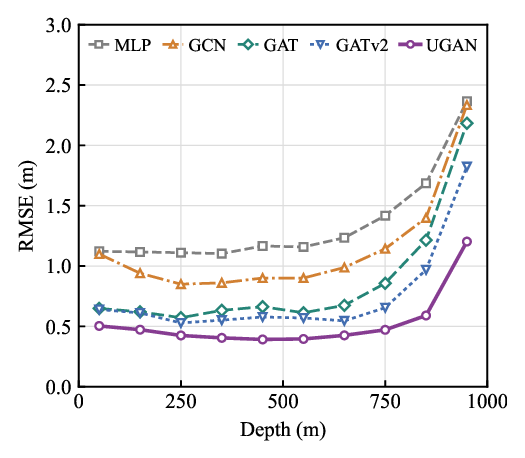}}\\
	\subfloat[$\gamma=0.010$]{\includegraphics[width=0.7\linewidth]{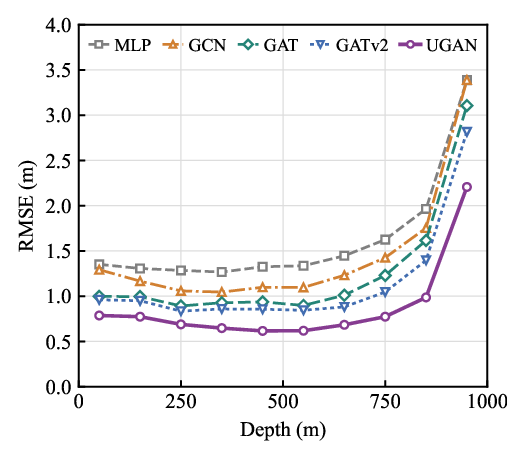}}
	\caption{Mean three-dimensional localization RMSE by target depth for the learning-based methods. Panels (a), (b), and (c) correspond to	$\gamma=0.002$, $0.005$, and $0.01$, respectively.}
	\label{fig:depth_rmse}
\end{figure}

Fig.~\ref{fig:cdf_p95} shows the empirical cumulative distribution functions (CDFs) of the Euclidean position errors for the learning-based methods. The empirical CDFs and P95 values are calculated from the test errors and reported as averages. UGAN yields the lowest P95 at each noise level. Its P95 values are $0.522$, $0.959$, and $1.638$~m, compared with $1.233$, $1.527$, and $2.219$~m for GATv2, respectively. The CDFs are displayed over $0$--$4$~m; this display range does not truncate the errors used to calculate the CDFs or P95 values.

Fig.~\ref{fig:depth_rmse} reports the three-dimensional RMSE in ten $100$-m intervals defined by the true target depth. RMSE is calculated within each interval and reported as an average. In the $900$--$1000$~m interval at $\gamma=0.005$, UGAN yields an RMSE of $1.203$~m, compared with $1.824$~m for GATv2. Across the three noise levels, the RMSE of UGAN in this interval rises from $0.548$ to $2.207$~m, indicating that near-seabed localization remains sensitive to TOA noise.

\begin{figure}[!t]
	\centering
	\captionsetup[subfloat]{font=footnotesize,labelfont=rm,textfont=rm}
\setcounter{subfigure}{0}
\subfloat[Horizontal error distribution]{\includegraphics[width=0.7\columnwidth]{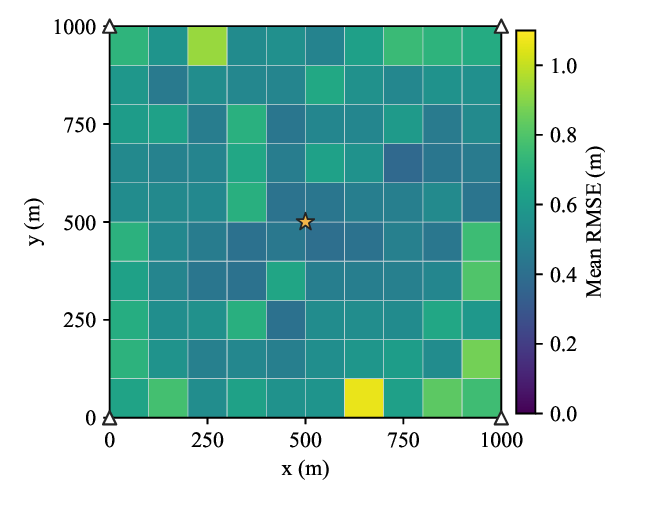}}
\par\vspace{0.5em}
\subfloat[Samples per spatial bin]{\includegraphics[width=0.7\columnwidth]{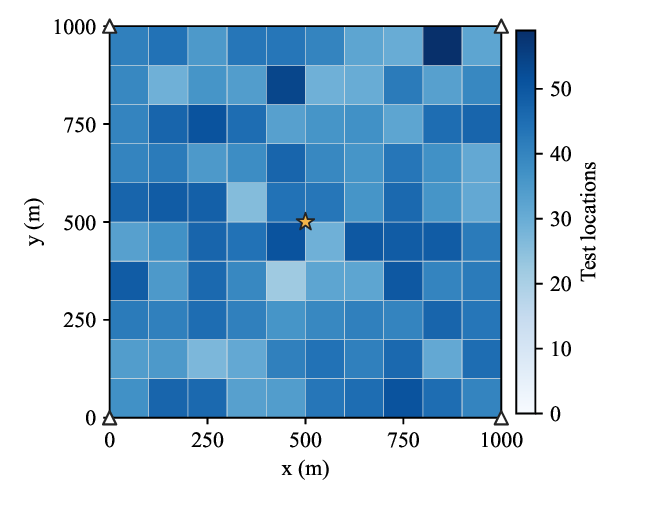}}
\par\vspace{0.4em}
\includegraphics[width=0.82\columnwidth]{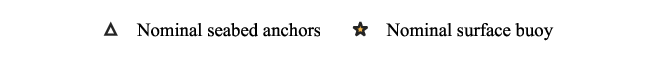}
	\caption{Horizontal distribution of UGAN localization error at $\gamma=0.005$. Panel (a) shows the mean three-dimensional RMSE in each bin. Panel (b) shows the number of test locations per bin. Triangles and the star mark the nominal horizontal positions of the seabed anchors and surface buoy, respectively.}
	\label{fig:spatial_rmse}
\end{figure}

Fig.~\ref{fig:spatial_rmse} shows the horizontal distribution of UGAN's three-dimensional localization RMSE at $\gamma=0.005$. Test samples are grouped by their true horizontal coordinates
into $100$~m $\times$ $100$~m bins. Each bin pools samples from all target depths. The three-dimensional RMSE is computed within each bin and reported as an average. The sample-count panel reports the number of test locations in each bin.

The mean bin RMSE ranges from $0.371$ to $1.064$~m, with $22$--$59$ test locations per bin. These values describe the horizontal variation in localization error for the fixed deployment and reflect the depth distribution of the test samples within each bin.

\subsection{Ablation Study}
At $\gamma=0.005$, the full UGAN model is compared with three variants to assess the effects of explicit geometry--time relation scoring, relation messages, and the individual TOA branch.

\textit{UGAN w/o Relation Scoring} removes the geometry--time relation term from the attention score in all three GTA layers. The attention scores are computed from the current node states. The first layer therefore scores messages using the initial coordinate features, and its unused relation encoder is removed. The second and third layers retain the relation messages, and their attention scores can still depend indirectly on timing and relation information encoded in the updated node states. The individual TOA branch remains active in all three layers.

\textit{UGAN w/o Relation Messages} removes the relation-message projections and the corresponding weighted relation contributions in the second and third GTA layers, which is equivalent to setting $\eta_2=\eta_3=0$ in \eqref{eq24}. The first layer already has $\eta_1=0$. Geometry--time relation encoding and explicit relation scoring remain active in all three layers, as do the shared individual TOA encoder and its layer-specific injection projections. This variant isolates the contribution of using relation features as message content when they are already used to determine attention weights.

\textit{UGAN w/o TOA Branch} removes the shared nonlinear encoder for individual TOA observations and the corresponding projections in all three GTA layers. As discussed in Remark~\ref{rem:toa-representations}, paired TOA observations remain available through relation scoring and relation messages in this variant. The comparison therefore quantifies the contribution of the individual TOA branch in the presence of paired TOA information.

\begin{figure}[!t]
\centering
\includegraphics[width=0.7\columnwidth]{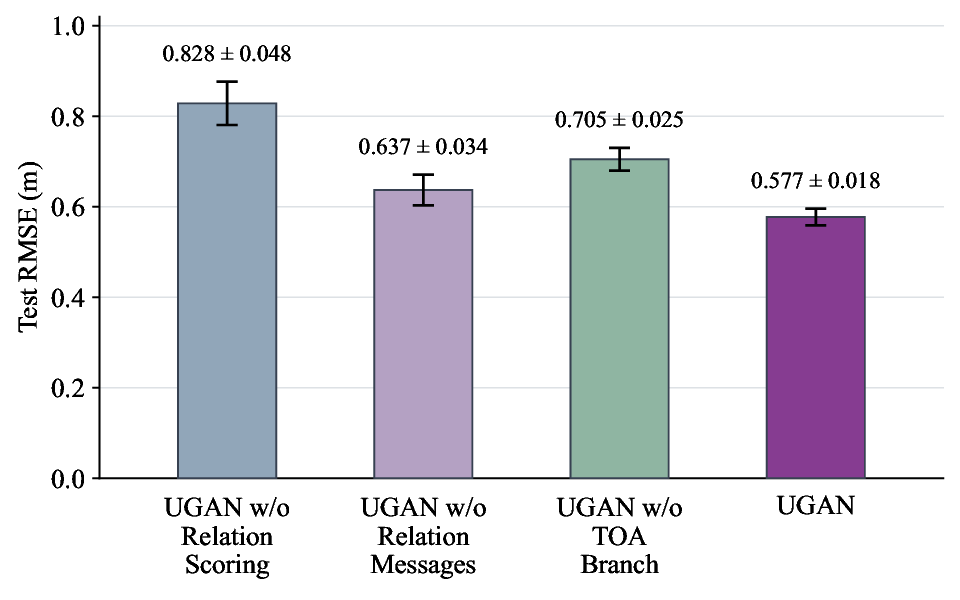}
\caption{Key-module ablation results at $\gamma=0.005$ for the full UGAN model and variants without relation scoring, relation messages, or the individual TOA branch. Bars and error bars show the mean and sample standard deviation of test RMSE, respectively.}
\label{fig:relation_ablation}
\end{figure}

As shown in Fig.~\ref{fig:relation_ablation}, the full UGAN model achieves a test RMSE of $0.577\pm0.018$~m. The RMSEs of UGAN w/o Relation Scoring, UGAN w/o Relation Messages, and UGAN w/o TOA Branch are $0.828\pm0.048$~m, $0.637\pm0.034$~m, and $0.705\pm0.025$~m, respectively. Relative to these variants, the full model reduces mean RMSE by $30.3\%$, $9.4\%$, and $18.2\%$, respectively. Removing relation messages increases mean RMSE by $0.060$~m even though explicit relation scoring is retained, supporting an additional benefit from incorporating relation features into the aggregated messages. Conversely, the higher RMSE without relation scoring shows that retaining relation messages alone does not recover the performance of the full relation mechanism. The individual TOA branch also provides an additional benefit when paired-TOA information remains available. These comparisons support the respective contributions of the three components within the evaluated architecture and noise setting; the TOA-branch ablation assesses that branch as a whole and does not separately establish the benefit of repeated injection over first-layer-only injection.

\subsection{Computational Cost and Execution Efficiency}
\label{sec:computational-cost}
Table~\ref{tab:complexity} reports the number of trainable parameters and multiply--accumulate operations (MACs) for the learning-based methods. MACs are counted for one forward pass on
a single five-station observation graph. The count includes linear projections, attention dot products, and graph aggregation, with activation functions, normalization, and softmax excluded. 
\begin{table}[!t]
	\caption{Model size and computational cost for single-target localization with five reference stations}
	\label{tab:complexity}
	\centering
	\papertableformat
	\begin{tabular*}{\columnwidth}{@{\extracolsep{\fill}}lcc@{}}
		\toprule
		Model & Parameters ($10^6$) & MACs per graph ($10^6$) \\
		\midrule
		MLP   & 3.173 &  3.169 \\
		GCN   & 3.157 & 15.829 \\
		GAT   & 2.248 & 10.745 \\
		GATv2 & 4.353 & 21.298 \\
		UGAN  & 3.684 & 18.392 \\
		\bottomrule
	\end{tabular*}
\end{table}

UGAN has $3.684$ million trainable parameters and requires $18.392$ million MACs per graph. Compared with the evaluated GATv2 configuration, UGAN uses $15.4\%$ fewer parameters and $13.6\%$
fewer MACs while achieving lower localization RMSE at all three TOA noise levels (Table~\ref{tab:performance}). MLP, GCN, and GAT have fewer parameters and require fewer MACs than UGAN.

To measure execution efficiency, we use the checkpoints trained with seed 42 at $\gamma=0.005$ and evaluate the five learning-based methods and constant-speed NLS on identical subsets of the same $4000$-sample test set. Each learning-based method retains its own training-set normalization statistics. A fixed seed-42 permutation defines nested subsets of $1$, $10$, $50$, $100$, $500$, $1000$, $2000$, and $4000$ targets, with five reference stations per target. The learning-based methods use a maximum inference batch size of $256$ on an NVIDIA GeForce RTX 5060 Laptop GPU with PyTorch 2.8.0 and CUDA 12.8, using single-precision inference, one PyTorch CPU thread, and deterministic operations with TF32 disabled. NLS uses MATLAB R2022b on an Intel Core Ultra 9 275HX processor, with one computational thread and sequential processing of targets. It retains the $1500$~m/s constant-speed model, nominal seabed-anchor coordinates, realized buoy coordinates, initialization, bounds, and stopping criteria of the accuracy baseline, and uses double-precision bounded trust-region-reflective least squares with an analytical Jacobian. For every method and target count, ten warm-up executions precede $30$ timed repetitions. The five GPU methods were measured in randomized order within each repetition; NLS was measured separately on the same computer using the saved sample order.

\begin{figure}[!t]
\centering
\includegraphics[width=0.90\columnwidth]{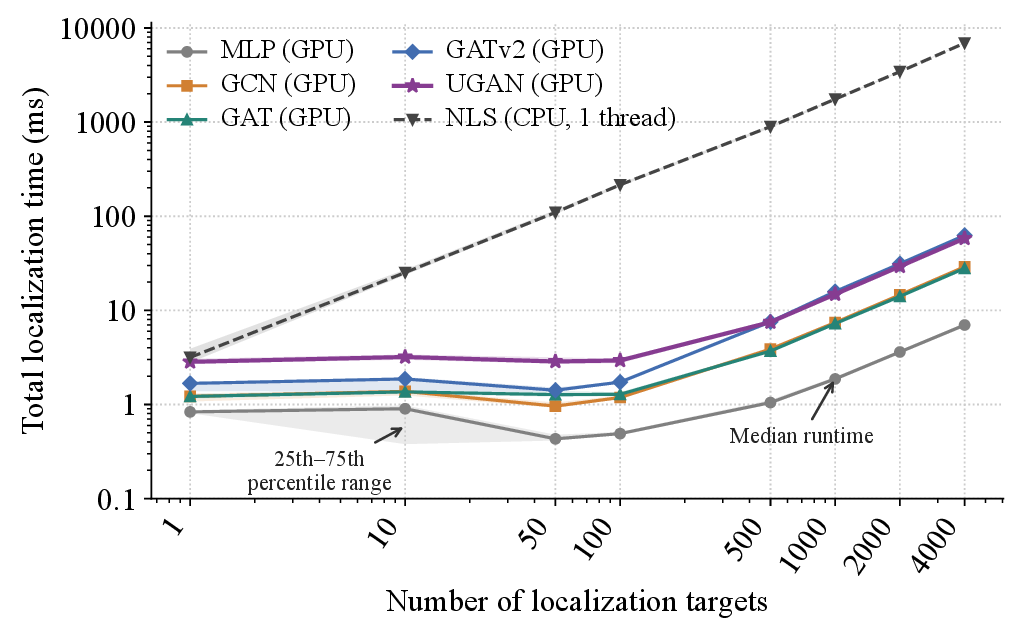}
\caption{Total localization time versus the number of targets for five learning-based methods (GPU) and constant-speed NLS (CPU, one computational thread), using identical test subsets at $\gamma=0.005$ and five reference stations per target. Curves show medians and shaded regions show interquartile ranges over 30 repetitions after warm-up. GPU inference uses a maximum batch size of 256; NLS solves targets sequentially. Both axes use logarithmic scales.}
\label{fig:execution_efficiency}
\end{figure}

The measured interval starts from raw TOA and station-coordinate arrays already in host memory and ends with the position estimates in meters returned to host memory. For the learning-based methods, it includes normalization, input-tensor assembly, host--device transfers, model execution including graph-feature construction, and coordinate restoration; CUDA is synchronized immediately before and after the timed interval. For NLS, it includes double-precision input preparation, output allocation, initial-guess construction, sequential iterative solution, and the original solver's diagnostic calculations. Program startup, model loading, disk access, sample selection, offline training, warm-up, error statistics, and plotting are excluded. Neither execution path performs ray tracing during localization. Fig.~\ref{fig:execution_efficiency} reports the median total time, with shaded regions indicating the interquartile range across repetitions.

At $4000$ targets, the median total times are $6.99$~ms for MLP, $29.02$~ms for GCN, $27.96$~ms for GAT, $62.51$~ms for GATv2, and $58.14$~ms for UGAN, compared with $6.91$~s for CPU single-thread NLS. For one target, NLS takes $3.16$~ms. These CPU/GPU results characterize the specified hardware, numerical precision, and implementations; their time differences cannot be attributed solely to algorithmic design. UGAN takes approximately $7.0\%$ less time than GATv2 at $4000$ targets, while MLP, GCN, and GAT are faster. This ordering is workload-dependent: for one target, UGAN takes $2.84$~ms compared with $1.68$~ms for GATv2. Thus, fewer MACs do not imply lower elapsed time at every target count. From $1000$ to $4000$ targets, the measured UGAN time grows approximately in proportion to the number of targets under the fixed maximum batch size. Small-workload timings are not strictly monotonic and are influenced by execution overhead and runtime variability. The $4000$-target time divided by $4000$ represents average processing cost per target in that workload, rather than the latency of an isolated target.

\section{Conclusion}
\label{sec:5}

This paper presented UGAN for three-dimensional underwater localization using multistation TOA observations and reference-station coordinates without requiring the current SSP at inference. UGAN constructs one observation graph per target and processes multiple graphs in a batch. Its Geometry--Time Attention mechanism uses station-pair geometry and ordered TOA pairs to compute attention weights in all three graph layers and to enrich the messages in the final two layers. Individual TOA features are injected directly into the node messages of each layer.

In ray-tracing simulations with a fixed five-station deployment and two held-out SSPs, UGAN achieved mean localization RMSEs of $0.307$, $0.577$, and $0.989$~m at the three evaluated TOA noise levels, corresponding to reductions of $52.0\%$, $31.0\%$, and $22.8\%$ relative to GATv2. At $\gamma=0.005$, the full model reduced mean RMSE by $30.3\%$, $9.4\%$, and $18.2\%$ relative to the variants without explicit relation scoring, relation messages, and the individual TOA branch, respectively. On the same GPU, UGAN processed 4,000 target observations in a median time of $58.14$~ms, compared with $62.51$~ms for GATv2, a reduction of $7.0\%$. Near-seabed localization remained sensitive to TOA noise. Future work will use field measurements to evaluate performance across a wider range of SSPs, station layouts, and anchor installation offsets.

\bibliographystyle{IEEEtran}
\bibliography{references}

@article{akyildiz,
  author = {I. F. Akyildiz and D. Pompili and T. Melodia},
  title = {{Underwater acoustic sensor networks: research challenges}},
  journal = {Ad Hoc Networks},
  volume = {3},
  number = {3},
  pages = {257--279},
  year = {2005},
  doi = {10.1016/j.adhoc.2005.01.004},
}

@article{tan,
  author = {H. P. Tan and R. Diamant and W. K. G. Seah and others},
  title = {{A survey of techniques and challenges in underwater localization}},
  journal = {Ocean Engineering},
  volume = {38},
  number = {14-15},
  pages = {1663--1676},
  year = {2011},
  doi = {10.1016/j.oceaneng.2011.07.017},
}

@article{erol,
  author = {M. Erol-Kantarci and H. T. Mouftah and S. Oktug},
  title = {{A Survey of Architectures and Localization Techniques for Underwater Acoustic Sensor Networks}},
  journal = {IEEE Communications Surveys \& Tutorials},
  volume = {13},
  number = {3},
  pages = {487--502},
  year = {2011},
  doi = {10.1109/surv.2011.020211.00035},
}

@article{paull,
  author = {L. Paull and S. Saeedi and M. Seto and others},
  title = {{AUV Navigation and Localization: A Review}},
  journal = {IEEE Journal of Oceanic Engineering},
  volume = {39},
  number = {1},
  pages = {131--149},
  year = {2014},
  doi = {10.1109/joe.2013.2278891},
}

@article{toky,
  author = {A. Toky and R. P. Singh and S. Das},
  title = {{Localization schemes for Underwater Acoustic Sensor Networks - A Review}},
  journal = {Computer Science Review},
  volume = {37},
  note = {Art. no. 100241},
  year = {2020},
  doi = {10.1016/j.cosrev.2020.100241},
}

@article{berger,
  author = {C. R. Berger and S. Zhou and P. Willett and others},
  title = {{Stratification Effect Compensation for Improved Underwater Acoustic Ranging}},
  journal = {IEEE Transactions on Signal Processing},
  volume = {56},
  number = {8},
  pages = {3779--3783},
  year = {2008},
  doi = {10.1109/tsp.2008.924801},
}

@article{rayls,
  author = {P. M. Ameer and L. Jacob},
  title = {{Localization Using Ray Tracing for Underwater Acoustic Sensor Networks}},
  journal = {IEEE Communications Letters},
  volume = {14},
  number = {10},
  pages = {930--932},
  year = {2010},
  doi = {10.1109/lcomm.2010.090810.101237},
}

@article{cheniso,
  author = {G. Chen and G. Wang and K. C. Ho and others},
  title = {{Underwater Acoustic Source Localization by TOA With Indeterminate Isogradient Sound Speed Profile}},
  journal = {IEEE Internet of Things Journal},
  volume = {12},
  number = {17},
  pages = {35522--35542},
  year = {2025},
  doi = {10.1109/jiot.2025.3580148},
}

@article{calibration,
  author = {L. Yang and K. C. Ho},
  title = {{Alleviating Sensor Position Error in Source Localization Using Calibration Emitters at Inaccurate Locations}},
  journal = {IEEE Transactions on Signal Processing},
  volume = {58},
  number = {1},
  pages = {67--83},
  year = {2010},
  doi = {10.1109/tsp.2009.2028947},
}

@article{geometry,
  author = {Y. Zhang and Y. Li and Y. Zhang and others},
  title = {{Underwater Anchor-AUV Localization Geometries With an Isogradient Sound Speed Profile: A CRLB-Based Optimality Analysis}},
  journal = {IEEE Transactions on Wireless Communications},
  volume = {17},
  number = {12},
  pages = {8228--8238},
  year = {2018},
  doi = {10.1109/twc.2018.2875432},
}

@article{bianco,
  author = {M. J. Bianco and P. Gerstoft and J. Traer and others},
  title = {{Machine learning in acoustics: Theory and applications}},
  journal = {The Journal of the Acoustical Society of America},
  volume = {146},
  number = {5},
  pages = {3590--3628},
  year = {2019},
  doi = {10.1121/1.5133944},
}

@article{niu,
  author = {H. Niu and E. Reeves and P. Gerstoft},
  title = {{Source localization in an ocean waveguide using supervised machine learning}},
  journal = {The Journal of the Acoustical Society of America},
  volume = {142},
  number = {3},
  pages = {1176--1188},
  year = {2017},
  doi = {10.1121/1.5000165},
}

@article{wangtransfer,
  author = {W. Wang and H. Ni and L. Su and others},
  title = {{Deep transfer learning for source ranging: Deep-sea experiment results}},
  journal = {The Journal of the Acoustical Society of America},
  volume = {146},
  number = {4},
  pages = {EL317--EL322},
  year = {2019},
  doi = {10.1121/1.5126923},
}

@article{chencnn,
  author = {R. Chen and H. Schmidt},
  title = {{Model-based convolutional neural network approach to underwater source-range estimation}},
  journal = {The Journal of the Acoustical Society of America},
  volume = {149},
  number = {1},
  pages = {405--420},
  year = {2021},
  doi = {10.1121/10.0003329},
}

@article{diniz,
  author = {P. Diniz and R. Calazan},
  title = {{Integrating modeled environmental variability into neural network training for underwater source localization}},
  journal = {The Journal of the Acoustical Society of America},
  volume = {153},
  number = {6},
  pages = {3201},
  year = {2023},
  doi = {10.1121/10.0019632},
}

@article{broad,
  author = {J. Yan and M. Yi and X. Yang and others},
  title = {{Broad-Learning-Based Localization for Underwater Sensor Networks With Stratification Compensation}},
  journal = {IEEE Internet of Things Journal},
  volume = {10},
  number = {15},
  pages = {13123--13137},
  year = {2023},
  doi = {10.1109/jiot.2023.3260192},
}

@article{bronstein,
  author = {M. M. Bronstein and J. Bruna and Y. LeCun and others},
  title = {{Geometric Deep Learning: Going beyond Euclidean data}},
  journal = {IEEE Signal Processing Magazine},
  volume = {34},
  number = {4},
  pages = {18--42},
  year = {2017},
  doi = {10.1109/msp.2017.2693418},
}

@article{wu,
  author = {Z. Wu and S. Pan and F. Chen and others},
  title = {{A Comprehensive Survey on Graph Neural Networks}},
  journal = {IEEE Transactions on Neural Networks and Learning Systems},
  volume = {32},
  number = {1},
  pages = {4--24},
  year = {2021},
  doi = {10.1109/tnnls.2020.2978386},
}

@inproceedings{gcn,
  author = {T. N. Kipf and M. Welling},
  title = {{Semi-supervised classification with graph convolutional networks}},
  booktitle = {International Conference on Learning Representations},
  year = {2017},
  url = {https://arxiv.org/abs/1609.02907},
}

@inproceedings{gat,
  author = {P. Veli\v{c}kovi\'{c} and G. Cucurull and A. Casanova and others},
  title = {{Graph attention networks}},
  booktitle = {International Conference on Learning Representations},
  year = {2018},
  url = {https://arxiv.org/abs/1710.10903},
}

@inproceedings{gatv2,
  author = {S. Brody and U. Alon and E. Yahav},
  title = {{How attentive are graph attention networks?}},
  booktitle = {International Conference on Learning Representations},
  year = {2022},
  url = {https://arxiv.org/abs/2105.14491},
}

@article{jointsync,
  author = {J. Liu and Z. Wang and J. H. Cui and others},
  title = {{A Joint Time Synchronization and Localization Design for Mobile Underwater Sensor Networks}},
  journal = {IEEE Transactions on Mobile Computing},
  volume = {15},
  number = {3},
  pages = {530--543},
  year = {2016},
  doi = {10.1109/tmc.2015.2410777},
}

@article{async,
  author = {J. Yan and X. Zhang and X. Luo and others},
  title = {{Asynchronous Localization With Mobility Prediction for Underwater Acoustic Sensor Networks}},
  journal = {IEEE Transactions on Vehicular Technology},
  volume = {67},
  number = {3},
  pages = {2543--2556},
  year = {2018},
  doi = {10.1109/tvt.2017.2764265},
}

@article{sspsimplify,
  author = {L. Bu and G. Li and X. Jia and others},
  title = {{Travel-time-weighted adaptive simplification of sound speed profiles for high-accuracy underwater acoustic ray tracing}},
  journal = {Ocean Engineering},
  volume = {352},
  note = {Art. no. 124682},
  year = {2026},
  doi = {10.1016/j.oceaneng.2026.124682},
}

@article{unified,
  author = {Y. Liu and Y. Wang and C. Chen and others},
  title = {{Unified Underwater Acoustic Localization and Sound Speed Estimation for an Isogradient Sound Speed Profile}},
  journal = {IEEE Sensors Journal},
  volume = {24},
  number = {3},
  pages = {3317--3327},
  year = {2024},
  doi = {10.1109/jsen.2023.3341459},
}

@article{ctlsl,
  author = {L. N. Liang and X. Mei and K. Zhang and others},
  title = {{CTLSL: A Robust Underwater Target Localization in Stratified and Perturbed UIoT Environments}},
  journal = {IEEE Internet of Things Journal},
  volume = {13},
  number = {19},
  pages = {44773--44785},
  year = {2026},
  doi = {10.1109/jiot.2026.3720542},
}

@article{limits,
  author = {Y. Shen and M. Z. Win},
  title = {{Fundamental Limits of Wideband Localization--- Part I: A General Framework}},
  journal = {IEEE Transactions on Information Theory},
  volume = {56},
  number = {10},
  pages = {4956--4980},
  year = {2010},
  doi = {10.1109/tit.2010.2060110},
}

@article{placement,
  author = {S. Xu and Y. Ou and X. Wu},
  title = {{Optimal Sensor Placement for 3-D Time-of-Arrival Target Localization}},
  journal = {IEEE Transactions on Signal Processing},
  volume = {67},
  number = {19},
  pages = {5018--5031},
  year = {2019},
  doi = {10.1109/tsp.2019.2932872},
}

@article{beacon,
  author = {M. Vemula and M. F. Bugallo and P. M. Djuri\'{c}},
  title = {{Sensor self-localization with beacon position uncertainty}},
  journal = {Signal Processing},
  volume = {89},
  number = {6},
  pages = {1144--1154},
  year = {2009},
  doi = {10.1016/j.sigpro.2008.12.019},
}

@article{mfp,
  author = {A. B. Baggeroer and W. A. Kuperman and P. N. Mikhalevsky},
  title = {{An overview of matched field methods in ocean acoustics}},
  journal = {IEEE Journal of Oceanic Engineering},
  volume = {18},
  number = {4},
  pages = {401--424},
  year = {1993},
  doi = {10.1109/48.262292},
}

@article{grnn,
  author = {Y. Wang and H. Peng},
  title = {{Underwater acoustic source localization using generalized regression neural network}},
  journal = {The Journal of the Acoustical Society of America},
  volume = {143},
  number = {4},
  pages = {2321--2331},
  year = {2018},
  doi = {10.1121/1.5032311},
}

@article{adapt,
  author = {D. Kari and H. Vishnu and A. C. Singer},
  title = {{Joint Source-Environment Adaptation of Data-Driven Underwater Acoustic Source Ranging Based on Model Uncertainty}},
  journal = {IEEE Journal of Oceanic Engineering},
  volume = {51},
  number = {2},
  pages = {1534--1549},
  year = {2026},
  doi = {10.1109/joe.2025.3638518},
}

@article{filters,
  author = {F. Gama and E. Isufi and G. Leus and others},
  title = {{Graphs, Convolutions, and Neural Networks: From Graph Filters to Graph Neural Networks}},
  journal = {IEEE Signal Processing Magazine},
  volume = {37},
  number = {6},
  pages = {128--138},
  year = {2020},
  doi = {10.1109/msp.2020.3016143},
}

@article{attsurvey,
  author = {C. Sun and C. Li and X. Lin and others},
  title = {{Attention-based graph neural networks: a survey}},
  journal = {Artificial Intelligence Review},
  volume = {56},
  number = {S2},
  pages = {2263--2310},
  year = {2023},
  doi = {10.1007/s10462-023-10577-2},
}

@article{multirel,
  author = {Z. Li and Y. Zhao and Y. Zhang and others},
  title = {{Multi-relational graph attention networks for knowledge graph completion}},
  journal = {Knowledge-Based Systems},
  volume = {251},
  note = {Art. no. 109262},
  year = {2022},
  doi = {10.1016/j.knosys.2022.109262},
}

@article{beckrls,
  author = {A. Beck and P. Stoica and J. Li},
  title = {{Exact and approximate solutions of source localization problems}},
  journal = {IEEE Transactions on Signal Processing},
  volume = {56},
  number = {5},
  pages = {1770--1778},
  year = {2008},
  doi = {10.1109/TSP.2007.909342},
}

\end{document}